\documentclass[twocolumn,times]{aastex63}
\usepackage[utf8]{inputenc}
\usepackage[normalem]{ulem}
\usepackage{amsmath}
 \usepackage{array,multirow,graphicx}

\begin{document}

\title{Microphysics of Circumgalactic Turbulence Probed by Fast Radio Bursts and Quasars}

\author[0000-0002-4941-5333]{Stella Koch Ocker}
\affiliation{Cahill Center for Astronomy and Astrophysics, California Institute of Technology, Pasadena, CA 91125, USA}
\affiliation{Observatories of the Carnegie Institution for Science, Pasadena, CA 91101, USA}
\author[0000-0002-8739-3163]{Mandy C. Chen}
\affiliation{Cahill Center for Astronomy and Astrophysics, California Institute of Technology, Pasadena, CA 91125, USA}
\affiliation{Observatories of the Carnegie Institution for Science, Pasadena, CA 91101, USA}
\author[0000-0002-1013-4657]{S. Peng Oh}
\affiliation{Department of Physics, University of California, Santa Barbara, CA 93106, USA}
\author[0000-0003-2635-4643]{Prateek Sharma}
\affiliation{Department of Physics, Indian Institute of Science, Bengaluru 560012, India}

\correspondingauthor{Stella Koch Ocker}
\email{socker@caltech.edu}

\begin{abstract}
   The circumgalactic medium (CGM) is poorly constrained at the sub-parsec scales relevant to turbulent energy dissipation and regulation of multi-phase structure. Fast radio bursts (FRBs) are sensitive to small-scale plasma density fluctuations, which can induce multipath propagation (scattering). The amount of scattering depends on the density fluctuation spectrum, including its amplitude $C_{\rm n}^2$, spectral index $\beta$, and dissipation scale $l_{\rm i}$. We use quasar observations of CGM turbulence at $\gtrsim$ pc scales to infer $C_{\rm n}^2$, finding it to be $10^{-16}\lesssim C_{\rm n}^2\lesssim 10^{-9}$~m$^{-20/3}$ for hot ($T>10^6$~K) gas and $10^{-8}\lesssim C_{\rm n}^2\lesssim 10^{-4}$~m$^{-20/3}$ for cool ($10^4\lesssim T\lesssim 10^5$~K) gas, depending on the gas sound speed and density. These values of $C_{\rm n}^2$ are much smaller than those inferred in the interstellar medium at similar physical scales. {The resulting scattering delays from the hot CGM are negligible ($\ll1$~$\mu$s at 1~GHz), but are more detectable from the cool gas as either radio pulse broadening or scintillation, depending on the observing frequency and sightline geometry}. Joint quasar-FRB observations of individual galaxies can yield lower limits on $l_{\rm i}$, even if the CGM is not a significant scattering site. {An initial comparison between quasar and FRB observations {(albeit for different systems)} suggests $l_{\rm i}\gtrsim750$~km in $\sim10^4$~K gas in order for the quasar and FRB constraints to be consistent.} If a foreground CGM is completely ruled out as a source of scattering along an FRB sightline then $l_{\rm i}$ may be comparable to the smallest cloud sizes ($\lesssim$ pc) inferred from photoionization modeling of quasar absorption lines.
\end{abstract}

\keywords{Radio bursts -- Circumgalactic medium -- Turbulence -- Scattering}

\section{Introduction}

The flow of gas between intergalactic and interstellar media through the circumgalactic medium (CGM) fuels galaxy formation and subsequent evolution. 
Turbulence is fundamental to a range of ongoing processes in the CGM: it moderates heating, provides pressure support, generates magnetic dynamos, yields density fluctuations that mediate cooling, and mixes dust and metals, leading to CGM enrichment \citep[for a review, see][]{faucher2023_review}. While there is abundant evidence that the CGM is multi-phase, based primarily on high areal covering fractions of cool/warm ($\sim 10^4 - 10^5$ K) gas detected in large samples of quasar absorption lines spanning a range of halo sizes and redshifts \citep{tumlinson2017,rudie2019}, the mass budget, origin, and regulation of this cooler gas by the surrounding hot ($\gtrsim10^6$ K) medium, which is itself poorly constrained, remain uncertain. Turbulence likely plays a key role in coupling and structuring these different phases \citep{ji19,fielding2020,tan21,gronke2022,mohapatra2022,mohapatra23}. 

Observational constraints from integral field spectroscopy (IFS) of emission around bright quasars \citep{chen2023,chen2024,chen2025} and nonthermal broadening of quasar absorption lines by foreground gas \citep{rudie2019,qu2022,hchen2023} indicate that velocity fluctuations in the CGM are broadly consistent with subsonic Kolmogorov turbulence on roughly pc to 10 kpc scales in $10^4 - 10^5$ K gas \citep{hchen2023} and on $\approx 1 - 100$ kpc scales in $\gtrsim 10^6$ K gas (indirectly inferred through cool nebular tracers; \citealt{chen2023,chen2024}), with evidence for injection on scales $\sim 10-100$ kpc depending on the system and its local environment \citep{chen2025}. However, CGM turbulence remains poorly constrained at the small ($\ll$ pc) spatial scales relevant to energy dissipation and the regulation of multi-phase structure \citep[e.g.][]{hummels2019,nelson2020,butsky2024b}.

Fast radio bursts (FRBs) are one of the few astrophysical probes with the potential to constrain microscale physics in the CGM \citep{2019MNRAS.483..971V,jow2024}. Like other compact radio sources (e.g. pulsars), FRBs undergo multipath propagation, or scattering, through density fluctuations along the line-of-sight (LOS), and the characteristics of the scattering, including its frequency dependence and impact on the burst intensity profile, are related to the density fluctuation power spectrum \citep[e.g.][]{cordes1986,rickett1990}. FRB intersections through foreground halos are common; in addition to all FRBs probing the Galactic halo and a significant fraction probing halos in the Local Volume \citep{connor2022_halos}, intermediate-redshift halo intersections grow with increasing FRB source redshift \citep{2013ApJ...776..125M}, such that {most} FRBs at $z\gtrsim0.5$ {are expected to} pass within twice the virial radius of multiple Milky Way-mass galaxies \citep{ocker2022b}. As we will discuss, FRB scattering generically probes density fluctuations at $\lesssim$ au scales in the CGM.

Placing meaningful constraints on the CGM from FRB scattering requires understanding the relative scattering contributions (or ``scattering budget'') of all media along the LOS, including host galaxies and the Galactic interstellar medium (ISM). For FRBs with precise scattering budgets, scattering appears to predominantly arise from {the ISMs of host galaxies and the Milky Way} \citep{ocker2022,cordes2022,sammons2022}, with observational limits on CGM scattering at the $\lesssim 10$s of $\mu$s level at 1 GHz \citep{2019Sci...366..231P,connor2020,ocker2021halo}. These scattering upper limits suggest that the scattering strength (and hence the strength of turbulence) is over an order of magnitude smaller in the CGM than in the Galactic ISM \citep{ocker2021halo}. However, scattering from the CGM may increase for FRBs with multiple foreground halos \citep[e.g.][]{faber24}, and if the strength of CGM turbulence evolves as a function of galaxy properties and/or cosmic time \citep{ocker2022b}. 

Theoretical predictions of CGM scattering are essential to the interpretation of FRB scattering budgets and their subsequent use as foreground probes. Previous theoretical examination of FRB scattering in the CGM has focused on simple analytic prescriptions for the \cite{2018MNRAS.473.5407M} ``misty'' CGM model \citep{2019MNRAS.483..971V,jow2024}, and predicted more scattering from the CGM than is actually observed \citep{ocker2021halo,jow2024,masribas2025}. In this paper, we leverage the growing number of observations of CGM turbulence in both emission and absorption to empirically predict how much scattering may occur in the CGM. While line emission and absorption probe disparate spatial scales from FRB radio wave scattering, it is possible that CGM turbulence extends to sub-parsec scales. 

Observations of plasma turbulence across a wide range of conditions, including in situ measurements of the solar wind \citep{alexandrova2008,alexandrova2012} and very local ISM \citep{2019NatAs...3..154L} and remote measurements of the diffuse ISM \citep{spanglergwinn90,rickett2009,geiger2024}, indicate that dissipation occurs near or below the plasma kinetic scales, as expected for magnetized (and anisotropic) plasma turbulence \citep[e.g.][]{lithwick2001}. X-ray observations of the intracluster medium similarly indicate that density fluctuations persist down to scales well below the collisional mean free path \citep{zhuravleva2019}. Chromatic scattering of pulsars and masers is consistent with a dissipation scale $\sim 100 - 1000$ km in the ionized ISM \citep{spanglergwinn90,rickett2009,geiger2024}, comparable to the ion inertial scale $\lambda_i = V_A/\Omega_i \approx 230\ {\rm km}(n_e/{\rm cm^{-3})^{-1/2}}$, where $V_A$ is the Alfv\'{e}n speed, $\Omega_i$ is the ion cyclotron frequency, and the latter equality is for an ionized hydrogen plasma with electron density $n_e$. In the CGM, $n_e \sim 10^{-4} - 10^{-2}$ cm$^{-3}$ implies ion inertial scales $\lambda_i \sim 10^3 - 10^4$ km. Even for a conservatively small density $n_e \sim 10^{-5}$ cm$^{-3}$, $\lambda_i \sim 10^5$ km. 

Characteristic values of the plasma kinetic scales for the hot and cool CGM are shown in Table~\ref{tab:microscales}, along with length scales relevant to FRB scattering. {We also show the Debye length, the absolute lower limit to the kinetic regime.} These kinetic scales relevant to dissipation are orders of magnitude smaller than scales considered in CGM simulations, which typically focus on hydrodynamic (non-magnetized) turbulence but without explicit dissipation. Moreover, for typical CGM temperatures and densities the dissipation scale of hydrodynamic turbulence is larger than the collisional mean free path (which is $\approx 5000$ au -- 1 pc; \citealt{spitzer1956,sarazin1986}). If magnetic turbulence is present in the CGM at sub-parsec scales, then FRBs may offer the only means of constraining it.

This paper makes empirical predictions of FRB scattering in the CGM, based on observational data at larger spatial scales. In doing so, we provide a methodology for contextualizing FRB scattering measurements and their physical interpretation in the landscape of other CGM probes, and we demonstrate how quasar and FRB observations may be used in tandem to constrain CGM microphysics. Section~\ref{sec:theory} describes the theoretical formalisms relating velocity and density fluctuation spectra, which are used to translate between observational constraints from quasars and FRBs summarized in Section~\ref{sec:obs}. The results of converting quasar-inferred velocity fluctuations into estimates of density fluctuations and subsequent FRB scattering are given in Section~\ref{sec:results}. Implications are discussed in Section~\ref{sec:discuss}, {and conclusions summarized in Section~\ref{sec:conc}}.

\begin{deluxetable}{c l C C}\label{tab:microscales}
\tablecaption{{Length Scales Relevant to FRB Scattering in the CGM}}
\tabletypesize{\footnotesize}
\tablehead{\colhead{} & \colhead{} & \colhead{Hot CGM} & \colhead{Cool CGM}}
\startdata
& Injection (Outer) Scale $l_{\rm o}$ & 10-100~{\rm kpc} & \gtrsim10~{\rm pc}~(?)\\ \hline
\parbox[t]{2mm}{\multirow{3}{*}{\rotatebox[origin=c]{90}{Scattering}}} & Multipath Scale $r_{\rm mp}$ & 10^{-4}-10^{-3}~{\rm pc} & 10^{-3}-0.1~{\rm pc}  \\
& Fresnel Scale $r_F$ & {\rm au} & {\rm au} \\
& Diffractive Scale $r_{\rm diff}$ & 0.2~{\rm au} & 10^5~{\rm km}\\ \hline
\parbox[t]{2mm}{\multirow{5}{*}{\rotatebox[origin=c]{90}{Dissipation ($l_{\rm i}$)}}} & Ion Inertial Length $\lambda_i$ & 2\times10^4~{\rm km} & 2300~{\rm km}\\
& Ion Gyroradius $r_i$ & 9000~{\rm km} & 900~{\rm km} \\
& Electron Inertial Length $\lambda_e$ & 500~{\rm km} & 50~{\rm km} \\
& Electron Gyroradius $r_e$ & 200~{\rm km} & 20~{\rm km} \\
& Debye Length $\lambda_D$ & 7~{\rm km} & 0.07~{\rm km} \\
\enddata
\tablecomments{Order-of-magnitude estimates of length scales relevant to FRB scattering in the hot CGM assume $n_e\sim10^{-4}$ cm$^{-3}$, $T\sim10^6$ K, and $B\sim\mu$G, and in the cool CGM $n_e\sim10^{-2}$ cm$^{-3}$, $T\sim10^4$ K, and $B\sim\mu$G. Length scales are defined as follows: The multipath scale is the diameter of the scattering disk at the scattering screen (Eq.~\ref{eq:rmp}); the Fresnel scale approximately divides refractive from diffractive optics (Eq.~\ref{eq:fresnel}); and the diffractive scale is the scale at which the rms phase perturbations are unity (Eq.~\ref{eq:rdiff}). Microscales relevant to turbulent dissipation are evaluated for an ionized hydrogen plasma based on the following definitions: $\lambda_i = V_A/\Omega_i$, $r_i = v_{\rm th,i}/\Omega_i$, $\lambda_e = V_A/\Omega_e$, $r_e = v_{\rm th,e}/\Omega_e$, and $\lambda_D = \sqrt{\epsilon_0k_BT/n_e}$, where $V_A$ is the Alfv\'{e}n speed, $\Omega_{i/e}$ is the ion/electron cyclotron frequency, $v_{\rm th,i/e}$ is the ion/electron thermal speed, $\epsilon_0$ is the permittivity of free space, and $k_B$ is the Boltzmann constant. {Our analysis makes scattering predictions under the assumption that density fluctuations dissipate near kinetic scales ($< r_{\rm diff}$); potential alternative dissipation scales are discussed in Section~\ref{sec:discuss}.}}
\end{deluxetable}

\section{Theory Relating Density and Velocity Fluctuation Spectra}\label{sec:theory}

In the inertial range of a turbulent cascade, both velocity and density fluctuations follow power-law wavenumber spectra of the form:
\begin{align*}
    P_{\delta v}(q) &= C_v^2 q^{-\beta}, \\
    P_{\delta n_e}(q) &= C_{\rm n}^2 q^{-\beta}, \\
    (q_{\rm o} < q &< q_{\rm i})
\end{align*}
where $C_v^2$ and $C_{\rm n}^2$ are the spectral amplitudes, the wavenumber $q = 2\pi/l$ for a length scale $l$, and the spectra extend over an inertial range of wavenumbers between an outer scale $l_{\rm o} = 2\pi/q_{\rm o}$ and an inner scale $l_{\rm i} = 2\pi/q_{\rm i}$. For an inner scale {$l_{\rm i} \ll l$}, a length scale of interest, the density fluctuation variance at the scale {$l$} can be found by integrating $P_{\delta n_e}$, which gives
\begin{equation}\label{eq:dne-Cn2}
\begin{split}
    \langle \delta n_e^2 \rangle &= \frac{2(2\pi)^{4-\beta}C_{\rm n}^2 l^{\beta-3}}{\beta -3} \hspace{15pt} (\beta > 3, l \gg l_{\rm i}) \\
                                &\approx 3(2\pi)^{1/3} C_{\rm n}^2 l^{2/3} \hspace{15pt} (\beta = 11/3, l \gg l_{\rm i})
\end{split}
\end{equation}
where the second approximation is for a Kolmogorov spectral index. The velocity fluctuation variance has a similar relation to $C_v^2$. For $\beta = 11/3$, $\langle \delta n_e^2 \rangle$ in cm$^{-6}$, and $l$ in au, 
\begin{equation}\label{eq:Cn2-1}
    C_{\rm n}^2 \approx 6.41 \times 10^3\ {\rm m}^{-20/3}\ \frac{\langle \delta n_e^2 \rangle}{\rm cm^{-6}}\ \bigg(\frac{l}{\rm au}\bigg)^{-2/3}.
\end{equation}

\noindent The motivation for $l\sim$~au comes from the very small Fresnel scales below which diffractive scattering occurs, as we discuss in \S\ref{sec:obs-FRB}. We relate the density and velocity fluctuation spectra assuming that the density fluctuations are a passive scalar, and we focus on the subsonic regime relevant to the CGM (Mach numbers $\mathcal{M}\lesssim1$). The hot ($T\gtrsim10^6$ K) gas is treated separately from the cool ($10^4 \lesssim T \lesssim 10^5$ K) gas, as follows.

\subsection{Hot CGM}

For the hot CGM, adiabatic simulations of isotropic, homogeneous, subsonic turbulence indicate that the root-mean-square (rms) density fluctuations are related to the Mach number squared \citep{mohapatra2019,mohapatra2022},
\begin{equation}\label{eq:Mach}
    \delta n_e/\langle n_e \rangle = b \mathcal{M}^2,
\end{equation}
where $\langle n_e \rangle$ is the average, volume-weighted hot gas density and $b \approx 0.3 - 1$ (hereafter we assume $b \approx 1$; \citealt{federrath2010,konstandin2012}). The Mach number can be constrained from measured velocity structure functions (VSFs) via comparison of VSF slopes across different VSF orders, given an assumed sound speed (\citealt{chen2023,chen2024}; for analogous simulation analysis, see \citealt{mohapatra2022b}). For a Kolmogorov spectrum, the rms density fluctuations have the same power-law index in length scale $l$ as the rms velocity fluctuations, giving 
\begin{equation}\label{eq:dne-Mach}
    \frac{\delta n_e}{\langle n_e \rangle } \approx 3\times10^{-5}\  \bigg(\frac{\mathcal{M}}{0.3}\bigg)^2 \bigg(\frac{l}{\rm au}\bigg)^{1/3} \bigg(\frac{\rm 100\ kpc}{l_{\rm o}}\bigg)^{1/3},
\end{equation}
where we have chosen values for $\mathcal{M}$ and $l_{\rm o}$ representative of VSFs measured in the CGM \citep{chen2024}. The density fluctuation amplitude $C_{\rm n}^2$ can thus be expressed in terms of $\mathcal{M}$ as 
\begin{equation}\label{eq:Cn2-2}
    C_{\rm n}^2 \approx 10^{-12} \ {\rm m}^{-20/3} \times \bigg(\frac{\langle n_e \rangle}{10^{-4}\ {\rm cm^{-3}}}\bigg)^2 \bigg(\frac{\mathcal{M}}{0.3}\bigg)^4 \bigg(\frac{\rm 100\ kpc}{l_{\rm o}}\bigg)^{2/3},
\end{equation}
where we have {combined Equations~\ref{eq:Cn2-1} and \ref{eq:dne-Mach}} and scaled to densities typical of the hot CGM. This $C_{\rm n}^2 \sim 10^{-12} \ {\rm m}^{-20/3}$ is many orders of magnitude smaller than that typical in the Galactic ionized ISM, for which $\langle n_e \rangle \sim 0.1$ cm$^{-3}$, $\mathcal{M} \sim 1$, and $l_{\rm o} \sim 100$ pc yields $C_{\rm n}^2 \sim 10^{-3}$ m$^{-20/3}$ (directly comparable to values inferred from observations of pulsar scattering and DM variations; \citealt{armstrong95,krishnakumar15,ocker_bowshocks}). Hence, the contribution of the hot CGM to scattering is expected to be negligible, {even when the geometric amplification of scattering is large (as will be shown in Section~\ref{sec:results}).} Note, that for $\mathcal{M} > 1$, Equation~\ref{eq:Mach} should be modified to scale linearly with $\mathcal{M}$ \citep{mohapatra2019}. A linear scaling between density fluctuations and Mach number also holds if the hot CGM is highly stratified with strong buoyant restoring forces \citep{zhuravleva14b,mohapatra21}. Even with these modifications, the hot CGM will still be orders of magnitude away from producing a detectable signal.

\subsection{Cool CGM}

For the cool CGM, density fluctuations are expected to behave similarly to those in an isothermal turbulence cascade in which $\delta n_e$ scales linearly with $\mathcal{M}$ \citep{padoan97,konstandin2012,federrath2021}. The amplitudes of the velocity and density fluctuation spectra, $C_v^2$ and $C_{\rm n}^2$, can be related to each other by expressing $\mathcal{M}$ in terms of the rms velocity fluctuations, which are related to $C_v^2$ in a form analogous to Equation~\ref{eq:Cn2-1}, yielding
\begin{equation}\label{eq:Cv-Cn}
    C_v^2 = \bigg(\frac{c_s}{b\langle n_e \rangle}\bigg)^2C_{\rm n}^2
\end{equation}
where $b$ is the same constant as in Equation~\ref{eq:Mach}, $\langle n_e \rangle$ is the average cool gas electron density, and $c_s$ is the sound speed \citep{2021arXiv210711334S}. {For $c_s$, $\langle n_e \rangle$, and $C_{v}^2$ in their typical units,}
\begin{multline}\label{eq:Cv-Cn2}
    C_{\rm n}^2 \approx 10^{-6}\ {\rm m}^{-20/3} \bigg(\frac{C_v^2}{10^{-9}\ {\rm km^{4/3}\ s^{-2}}}\bigg) \\ \times \bigg(\frac{\langle n_e \rangle}{10^{-2}\ \rm cm^{-3}}\bigg)^{2} \bigg(\frac{c_s}{30\ \rm km\ s^{-1}}\bigg)^{-2}.
\end{multline}
The velocity spectral amplitude $C_v^2$ is constrained by measurements of nonthermal line widths, which can be used to infer $\langle \delta v^2 \rangle$. For nonthermal broadening dominated by turbulence, the nonthermal line width $b_{\rm NT}$ is related to the rms velocity fluctuations $\delta v$ as $\delta v = \sqrt{3/2}b_{\rm NT}$, and $\delta v \propto l^{1/3}$ for $\beta = 11/3$ \citep{hchen2023}. {Given $\delta v$ at a specific length scale $l$, $C_v^2$ can be derived analogously to Equation~\ref{eq:dne-Cn2}:}
\begin{equation}\label{eq:deltav-Cv2}
\begin{split}
    \langle \delta v^2 \rangle &\approx 3(2\pi)^{1/3} C_v^2 l^{2/3} \\
    &\approx 600\ {\rm km^2\ s^{-2}} \bigg(\frac{C_v^2}{\rm 10^{-9}~ km^{4/3}\ s^{-2}}\bigg) \bigg(\frac{l}{\rm kpc}\bigg)^{2/3}.
\end{split}
\end{equation}
For $\delta v \approx 20$ km/s at $l = 1$ kpc, $C_v^2 \approx 10^{-9}$ km$^{4/3}$ s$^{-2}$, the fiducial value adopted in Equation~\ref{eq:Cv-Cn2}. 
Note that inferring $C_{\rm n}^2$ from velocity fluctuations always requires an assumption of the sound speed, which is implicitly used to constrain $\mathcal{M}$ from measured VSFs. Moving forward, we assume that the density-Mach relations used above also apply to the volume-averaged density within a single phase.

\section{CGM Turbulence Observables}\label{sec:obs}

Observations of CGM turbulence are primarily at $\gg$ pc scales, from IFS of emission around bright quasars \citep{chen2023,chen2024,chen2025} and nonthermal broadening of quasar absorption lines \citep{rudie2019,qu2022,hchen2023}. Here we summarize key observational results from emission and absorption studies, followed by the FRB scattering observables that will be estimated and interpreted.

\subsection{Emission and Absorption}

IFS yields spatially resolved maps of emission line radial velocities, which constrain velocity fluctuations over spatial scales of $\approx 1 - 100$ kpc \citep{chen2023,chen2024,chen2025}. Nonthermal broadening of quasar absorption lines yields the sightline-integrated velocity dispersion, but associating the resulting velocity fluctuation amplitude with a specific spatial scale requires photoionization modeling of sometimes degenerate line components in order to identify discrete clouds and estimate their sizes \citep{hchen2023}. These two techniques have been applied to galaxy samples spanning different redshifts, masses, and active galactic nuclei (AGN) characteristics. 

We make use of three key results from the quasar studies described above:

\begin{enumerate}
    \item CGM turbulence is broadly consistent with a Kolmogorov spectral index\footnote{While deviations from Kolmogorov turbulence are seen in some systems, it is unclear whether these deviations are bona fide and physical \citep{chen2024}.} \citep{chen2023,hchen2023,chen2024}. We thus assume $\beta = 11/3$ for both hot and cool gas in subsequent analysis, and discuss the effect of modifying $\beta$ on our results.
    \item In emission, the velocity fluctuations are consistent with Mach numbers $\approx 0.2 - 1.8$. These velocities are inferred from emission by cool $T \sim 10^4$ K nebular gas, and are thought to reflect the motions of cool clumps entrained in hot  $T >10^6$ K gas \citep{chen2023,chen2024}. {The Mach number is inferred by comparing the velocity dispersion in the plane of the sky, $\sigma_{\rm pos}$, and the hot gas sound speed, and is corroborated by the VSF slopes.} Given that the majority of systems appear to be subsonic, we restrict our analysis to Mach numbers $\leq1$, {and we focus on densities typical of the hot CGM in Milky Way-mass galaxies, $10^{-5} \leq n_e \leq 10^{-3}$ cm$^{-3}$ \citep[e.g.][]{sharma2012,miller_bregman_2015,singh2018,voit2019}. While the hot CGM may have even higher densities in the core of the halo, we are primarily interested in FRB LOSs at $\sim10-200$ kpc from the halo center, where observations are not contaminated by the ISM.}
    \item In absorption, nonthermal line broadening indicates a roughly Kolmogorov scaling relation between cloud size and nonthermal line width for $10^4 - 10^5$ K gas, where the cloud sizes probed are $\sim$ pc -- 10 kpc, and the corresponding {local/physical cloud gas} densities are $n_e \sim  10^{-4} - 0.1$ cm$^{-3}$ {(the volume-averaged density $\bar{n}_e  = fn_e$ in the CGM is $\sim 10^{-3} - 10^{-2}$ times smaller for expected volume filling fractions of the cool gas \citep[e.g.][]{augustin2025}, whereas for the hot gas $f\sim1$ and $\bar{n}_e \approx n_e$)}. The fitted scaling relation gives the rms velocity fluctuations at 1 kpc, $\delta v(1\ {\rm kpc}) \approx 22$ km/s \citep{hchen2023}. The uncertainty in this anchor value of $\delta v$ has a miniscule effect on resulting estimates of $C_v^2$ when accounting for a range of possible densities and sound speeds, and so we leave $\delta v$ at 1 kpc fixed to 22 km s$^{-1}$ in subsequent analysis of the cool gas. Nonthermal broadening also yields estimates of the Mach number, which appears to be $\mathcal{M}<1$ for a majority of absorbers \citep{rudie2019,qu2022}.
\end{enumerate}

Based on the above results, we use the velocity fluctuations constrained in emission to estimate density fluctuations in the hot CGM, and fluctuations constrained in absorption for the cool CGM. {Applying these results to our scaling relations between $C_v^2$ and $C_{\rm n}^2$ (Section~\ref{sec:theory}) assumes a one-to-one correspondence for density and velocity fluctuations across all scales, which may not necessarily be the case.} We also note that these emission and absorption measurements trace different galaxy systems that are heterogeneous in mass, redshift, and AGN properties. There is initial evidence that turbulence may be more dominant in the CGM of quiescent galaxies than in star-forming galaxies and at $z\leq1$, although the physical interpretation of this trend is unclear and larger samples are needed for a robust, uniform comparison \citep{qu2022}. The inferred energy transfer rates per unit mass also appear to be $>10\times$ larger in the CGM around luminous quasars than in typical $L_*$ galaxies \citep{chen2023,chen2025}.
This difference may be related to both the differing gas phases probed and a more systematic difference in the galaxy properties like AGN activity. In future, direct comparisons between emission and/or absorption and FRB observations will preferably be performed within the same galaxy system.

\subsection{FRB Scattering}\label{sec:obs-FRB}

A continuous spectrum of electron density fluctuations will produce a range of refractive and diffractive effects that are observable on different (and sometimes overlapping) time and radio frequency scales \citep{cordes1986}. Refraction can include angle of arrival variations or image wandering, long-term (days to months) intensity modulations, and under certain conditions multiple imaging. Diffraction includes short-term (minutes to hours) intensity modulations, pulse broadening, and angular broadening. We use the term ``scintillation'' to refer to the generic phenomenon of chromatic intensity modulations caused by constructive and destructive interference of scattered radiation, which can be both refractive and diffractive in nature \citep{rickett1990}. The relevance of refractive and diffractive scattering is a function of spatial scale that depends on the shape, amplitude, and cutoff of the density fluctuation spectrum \citep{goodman85,cordes1986,coles87}. 

In general, diffractive effects dominate for Kolmogorov media when the inner scale is much smaller than the Fresnel scale, whereas if density fluctuations decline more strongly at small scales (i.e., the power spectrum has a steeper power law scaling), refraction can dominate across a wide range of spatial scales. Modifications to the density fluctuation spectrum (e.g. local enhancements or shocks and energy injection at multiple scales) also affect the resulting radio scattering observables. For simplicity, we focus subsequent analysis on a statistically homogeneous medium with a single injection scale, and wherever a fiducial assumption of the spectral index $\beta$ is needed we adopt $\beta = 11/3$. However, in principle the scattering observables treated here can be derived for a medium of arbitrary spectral index (see Appendix and for an in-depth treatment, \citealt{cordes1986}). 

\subsubsection{Length Scales of Diffractive Scattering}

Given that the majority of FRBs are only observed once, we focus on diffractive scattering effects that are routinely constrained within individual bursts, namely pulse broadening and scintillation. In the regime of diffractive scattering, radio waves undergo instantaneous multipath propagation, and the underlying plasma density fluctuations must extend to length scales significantly smaller than the Fresnel scale, $r_F$, which roughly separates the regimes of diffractive and refractive optics (see, however, \citealt{jow2023b}, for an in-depth treatment that accounts for the effect of lens strength in the refractive regime). 

\begin{figure}
    \centering
    \includegraphics[width=0.48\textwidth]{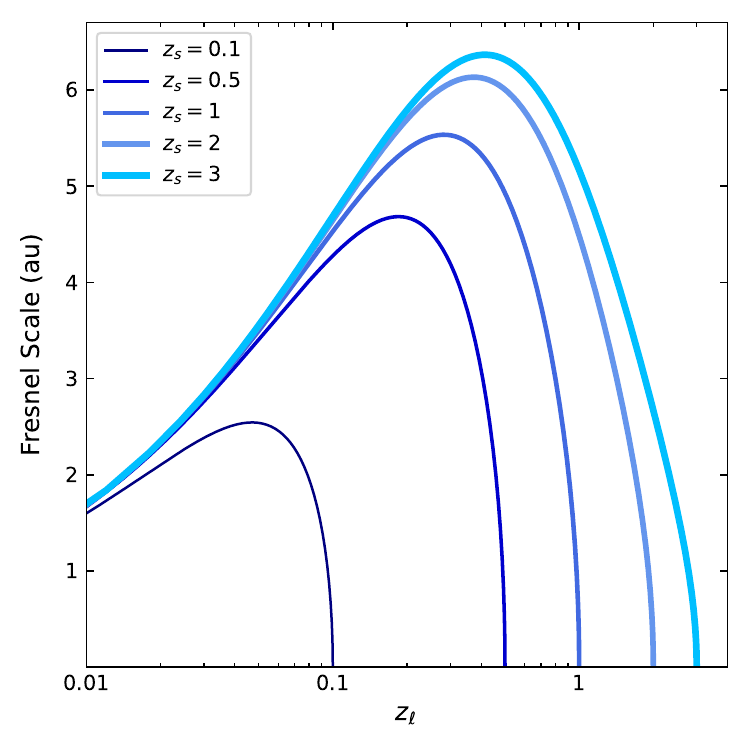}
    \caption{Fresnel scale vs. lens redshift ($z_\ell$) for a range of FRB source redshifts ($z_s$). The Fresnel scale is evaluated using angular diameter distances (Equation~\ref{eq:fresnel}). The diffractive scattering effects considered here arise from density fluctuations at length scales significantly smaller than the Fresnel scale.}
    \label{fig:fresnel-scale}
\end{figure}

For a scattering medium (or screen/lens) at a cosmological redshift $z_\ell$, the Fresnel scale is 
\begin{equation}\label{eq:fresnel}
    r_F = \bigg(\frac{d_{\rm lo}d_{\rm sl}\lambda_{\rm obs}}{2\pi d_{\rm so}(1+z_\ell)}\bigg)^{1/2}
\end{equation}
where $d_{\rm lo}$, $d_{\rm sl}$, and $d_{\rm so}$ are angular diameter distances between the lens and observer, source and lens, and source and observer, respectively, and the observing wavelength $\lambda_{\rm obs}$ is a factor $(1+z_\ell)$ smaller in the lens frame \citep{2013ApJ...776..125M}. Figure~\ref{fig:fresnel-scale} shows $r_F$ vs. $z_\ell$ for a range of source redshifts. For typical FRB redshifts ($z_s \lesssim 1$), $r_F$ is at most 2 to 5 au, and even for larger $z_s$ it does not exceed 10 au. The length scales relevant to diffractive scattering are thus orders of magnitude smaller than typical resolutions in numerical CGM simulations, and well below the spatial scales accessible to both absorption and emission line observations of the CGM. 

{The characteristic scattering angle, $\theta_{\rm d}$, is typically defined as $\theta_{\rm d} = (d_{\rm sl}/d_{\rm so})\lambda_{\rm obs}/2\pi r_{\rm diff}$, where the diffractive scale $r_{\rm diff}$ is the transverse separation in the observing plane across which there is an rms phase difference of 1 radian (equivalently, $r_{\rm diff}$ is the length scale for which the phase structure function is unity; see \citealt{rickett1990} for a review). For plane waves (as appropriate for FRB sources at cosmological distances) and the regime where $r_{\rm diff} > l_{\rm i}$, 
\begin{equation}\label{eq:rdiff}
    r_{\rm diff} = \bigg[\frac{2^{2-\beta} \pi^2 r_e^2 \lambda_{\rm obs}^2 \beta {\rm SM}}{(1+z_\ell)^2} \frac{\Gamma(-\beta/2)}{\Gamma(\beta/2)} \bigg]^{1/(2-\beta)},
\end{equation}
where $r_e$ is the classical electron radius, $\lambda_{\rm obs}$ is the observing wavelength ($\lambda_{\rm obs} \approx 0.3$ m for $\nu = 1$ GHz), and ${\rm SM} \approx C_{\rm n}^2 L$ for a path length $L$ through a screen of constant $C_{\rm n}^2$ \citep{rickett84,cordes1986,coles87}}\footnote{Note that Equation~\ref{eq:rdiff} is appropriate for wavenumber defined as $q = 2\pi/l$ and applies to $2 < \beta < 4$. \cite{cordes1986} also give expressions for $r_{\rm diff}$ (denoted $\Delta x_{d}$ in that paper) for $4 < \beta < 6$. As in \cite{cordes1986}, we define $r_{\rm diff}$ as the $1/e$ half-width of the visibility function. As in \cite{2013ApJ...776..125M}, we include a factor of $(1+z_\ell)^{-2}$ in  $r_{\rm diff}$, but we note that their expression for $r_{\rm diff}$ differs from ours by a factor of $\pi^{1/(2-\beta)}$.}. {The diameter of the scattering cone at the scattering screen, also known as the multipath scale, is then
\begin{equation}\label{eq:rmp}
r_{\rm mp} = 2d_{\rm lo}\theta_{\rm d} 
\end{equation}
\citep{cordes1986}. The multipath scale defines the maximum length scale of the density fluctuations driving multipath propagation. All of our scattering expressions apply in the so-called strong scattering regime, for which $r_{\rm mp} > r_F$. This regime applies for most of the conditions we are interested in; specific conditions under which this regime does not apply are discussed further in Section~\ref{sec:results}.}

{We will refer to the time delay between an undeflected ray and a ray scattered through angle $\theta_{\rm d}$} as the diffractive scattering time, $\tau_{\rm d}$, {which is not the same as the mean scattering time, $\langle \tau \rangle$, that will be used throughout the rest of the paper.} The diffractive scattering time is $\tau_{\rm d} \propto (d_{\rm so}d_{\rm lo}/d_{\rm sl})(\theta_{\rm d}^2/2c)$. For a Gaussian scattered image, $\theta_{\rm d}$ is the $1/e$ half-width of the image, and the corresponding time delay is the $1/e$ time of a scattered pulse that temporally decays as a one-sided exponential. This special case of a Gaussian scattered image is conventionally assumed when inferring scattering delays from observed pulse shapes. {The diffractive scale $r_{\rm diff}$, and the corresponding delay $\tau_{\rm d}$, will depend on the inner scale of the density fluctuation spectrum only when $r_{\rm diff}\ll l_{\rm i}$; otherwise, $r_{\rm diff}$ does not depend on $l_{\rm i}$ (as shown in Equation~\ref{eq:rdiff}).}

However, in a turbulent medium with $\beta < 4$ and an inner scale $l_{\rm i} \ll r_{\rm diff}$, the scattered image is not Gaussian and instead has an extended halo that decays more slowly than a Gaussian, because the density fluctuations at scales {$l \ll r_{\rm diff}$ lead to scattering angles $\theta \sim \lambda/l \gg \lambda/r_{\rm diff}$} \citep{lambert99}. In the time domain, the scattered pulse decays more slowly than an exponential, and the pulse has a significant amount of radiation at time delays $\gg \tau_{\rm d}$. The mean scattering time of this pulse, $\langle \tau \rangle$, will be greater than the diffractive scattering time $\tau_{\rm d}$. {{While the diffractive scattering time will have no dependence on $l_{\rm i}$ in this regime, the mean scattering time does depend on $l_{\rm i}$ due to the long tail in the pulse broadening function as a result of scattered flux at angles $>\theta_{\rm d}$ (as shown in Appendix Figure~\ref{fig:scattering_diagram}).}} Conversely, when $r_{\rm diff} \ll l_{\rm i}$ (or $\beta\geq4$), the scattered image becomes Gaussian, the extended halo is absent, and $\langle \tau \rangle = \tau_{\rm d}$ \citep{rickett2009,geiger2024}. 

In the CGM, inferred diffractive scattering times $\lesssim 10$s of $\mu$s imply $r_{\rm diff} \gtrsim 10^6$ km \citep{2019Sci...366..231P}. If $l_{\rm i}$ is comparable to plasma kinetic scales, then we are in the regime where $l_{\rm i} \ll r_{\rm diff}$. In this regime, the scattered pulse will deviate from a one-sided exponential and include contributions from scattering angles $> \theta_{\rm d}$ ({{i.e., length scales smaller than the diffractive scale}}). The relevant time delays $\langle \tau \rangle > \tau_{\rm d}$ hence yield information about density fluctuations extending down to the inner scale $l_{\rm i}$, {{thus motivating our focus on the mean scattering time for the rest of our analysis}}. 

\subsubsection{Generic Ionized Cloudlet Model}

We proceed by relating the mean scattering time to underlying density fluctuations using a generic ionized cloudlet model, in which the CGM is a statistically homogeneous medium consisting of ionized cloudlets with a volume filling factor $f$, internal cloudlet density {$n_{ec}$}, and internal density fluctuations following a Kolmogorov spectrum down to scales $l_i$ smaller than the diffractive scale. The scattering strength of the medium is then recast in terms of a composite density fluctuation parameter (see Appendix) 
\begin{equation}
\begin{split}
    \tilde{F} &= (\zeta \epsilon^2/f)l_{\rm o}^{3-\beta} l_{\rm i}^{\beta-4}\\
    &= (\zeta \epsilon^2/f)(l_{\rm o}^2 l_{\rm i})^{-1/3}\ {\rm for}\ \beta=11/3,
\end{split}
\end{equation}
where {$\zeta = \langle n_{ec}^2\rangle/\langle n_{ec} \rangle^2$ is the density variation between cloudlets and $\epsilon^2 = (\langle \delta n_{ec})^2\rangle/n_{ec}^2$} is the fractional density variance within a cloudlet \citep{cordes2016,ocker2021halo,cordes2022}. 

This formulation has the advantage of yielding a simple relation between the mean scattering time in the observer frame $\langle \tau \rangle_{\rm obs}$ and the dispersion measure (DM) contributions of a given medium along the LOS,
\begin{equation}\label{eq:tau-Ftilde}
    \langle \tau \rangle_{\rm obs} \approx 48.03\ {\rm ns}\  \tilde{F} G_{\rm scatt}{\rm DM_\ell}^2 \nu^{-4} (1+z_\ell)^{-3}
\end{equation}
where $\nu$ is the observing frequency, $z_\ell$ is the screen redshift, $\rm DM_\ell$ is the screen DM (due to the gas phase under consideration) in its rest frame, and $G_{\rm scatt}$ is a geometric leverage factor that depends on the relative locations of the source, screen, and observer. For cosmological screens far from the observer and source, 
\begin{equation}\label{eq:Gscatt}
    G_{\rm scatt} = 2d_{\rm sl}d_{\rm lo}/Ld_{\rm so},
\end{equation}
for angular diameter distances defined as in Equation~\ref{eq:fresnel} and a path length through the screen $L$. The equivalent scintillation bandwidth, {defined as the frequency width of the intensity autocorrelation function}, is $\Delta \nu_{\rm d} = C/(2\pi\langle\tau\rangle)$, where $C$ is a constant of order unity depending on the density fluctuation spectrum; for a uniform Kolmogorov medium filling the entire LOS, $C = 1.16$, whereas for a Kolmogorov thin screen $C = 0.96$ \citep{cordesrickett98}. We adopt $C = 1$ in subsequent analysis, which has a negligible effect on our results.

Given that the majority of scattering times reported in the literature are $1/e$ times that assume a Gaussian scattered image, we require a means of converting the $1/e$ delay to the mean scattering time given by Equation~\ref{eq:tau-Ftilde}. This conversion is accomplished through the dimensionless factor $A_\tau \equiv \tau_{\rm d}/\langle \tau \rangle$, which directly depends on the ratio $l_{\rm i}/r_{\rm diff}$ {(and therefore implicitly on the observing frequency, which changes the strength of scattering and hence $r_{\rm diff}$)}. For a Kolmogorov spectrum and $\nu = 1$ GHz, $A_\tau \approx 1/6$ when $l_{\rm i}/r_{\rm diff} \ll 0.1$ {(with little variation below this value)}, and $A_\tau \approx 0.7$ for $l_{\rm i}/r_{\rm diff} = 1$ \citep{cordes2022}. In the CGM, $l_{\rm i} \sim 10^3 - 10^4$ km and $r_{\rm diff} \gtrsim 10^6$ km implying $l_{\rm i}/r_{\rm diff} \lesssim 10^{-3} - 10^{-2}$. We therefore adopt $A_\tau = 1/6$ when converting the $1/e$ delay $\tau_{\rm d}$ to the mean scattering time, and all $1/e$ delays are scaled to 1 GHz assuming $\tau_{\rm d} \propto \nu^{-4.4}$, as appropriate for $\beta = 11/3$. The mean scattering delay scales exactly as $\nu^{-4}$, but $A_\tau$ depends on $r_{\rm diff}$, and hence $\nu$, such that $\tau_{\rm d} \propto A_\tau \nu^{-4} \propto \nu^{-4.4}$, as expected. {In principle, the mean delay and deviations from a one-sided exponential can be directly inferred from pulse shapes, but this will likely only be possible for bright bursts without complex multi-component structure.}

For a volume-averaged density {$\bar{n}_e = f \langle n_{ec} \rangle$}, $\tilde{F}$ is related to {$\overline{C_{\rm n}^2} \equiv fC_{\rm n}^2$} as
\begin{equation}\label{eq:Ftilde-Cn2}
    \tilde{F} \approx 0.5\ ({\rm pc^2 \ km})^{-1/3}\ \bigg(\frac{\overline{C_{\rm n}^2}}{\rm m^{-20/3}}\bigg)\bigg(\frac{\bar{n}_e}{\rm cm^{-3}}\bigg)^{-2} \bigg(\frac{l_{\rm i}}{\rm km}\bigg)^{-1/3}
\end{equation}
(see Appendix~\ref{app}). Formally the dependence on $\bar{n}_e$ cancels out when evaluating $\overline{C_{\rm n}^2}$, which is proportional to $\bar{n}_e^2$ (Equation~\ref{eq:Cn2-2}). {Given uncertainties in the volume filling factor of cool CGM gas, we also define a local fluctuation parameter $\tilde{F}_l \equiv f\times\tilde{F} = C_{\rm n}^2 n_e^{-2} l_{\rm i}^{-1/3}$ (where $C_{\rm n}^2$ and $n_e$ again refer to local quantities). This local $\tilde{F}_l$ can be estimated purely from local $C_{\rm n}^2$ and its relation to the velocity fluctuation spectrum (Section~\ref{sec:theory}), without making any assumption about the underlying density fluctuations described by $f$, $\zeta$, and $\epsilon^2$. Such a formulation is useful because quasar absorption lines typically trace individual cool gas structures and hence yield estimates of local density and $C_{\rm n}^2$.}
 
For typical Galactic pulsar sightlines probing the diffuse ISM, $\overline{C_{\rm n}^2} \approx 10^{-3.5}$ m$^{-20/3}$, $l_{\rm i} \approx 1000\ {\rm km}$, and $\bar{n}_e\approx 0.02$ cm$^{-3}$, which yields $\tilde{F} \approx 0.04$ (pc$^2$ km)$^{-1/3}$. This estimate is entirely consistent with values of $\tilde{F}$ inferred directly from pulse broadening and DM measurements in the disk \citep{ocker2021halo}. In the hot CGM, densities $\sim 10^{-4}$ cm$^{-3}$ {and a filling factor $f\approx1$} yield $\overline{C_{\rm n}^2} \sim 10^{-12}$ m$^{-20/3}$ (Equation~\ref{eq:Cn2-2}), equivalent to $\tilde{F} \sim 10^{-5.3}$ (pc$^2$ km)$^{-1/3}$ for $l_{\rm i} \sim 1000$ km. {For the cool CGM, typical local values of $C_{\rm n}^2 \sim 10^{-6}$ m$^{-20/3}$ and $n_e\sim10^{-2}$ cm$^{-3}$ would yield $\tilde{F}_l \sim 10^{-3}$ (pc$^2$ km)$^{-1/3}$ for $l_{\rm i}\sim1000$ km. Within a single cool cloud, $f\sim1$ and $\tilde{F} = \tilde{F}_l$; for the entire halo, $f\ll 1$ and $\tilde{F}$ will be larger, depending on the assumed filling factor.}

\section{Analysis \& Results}\label{sec:results}

\begin{figure*}
    \centering
    \includegraphics[width=\textwidth]{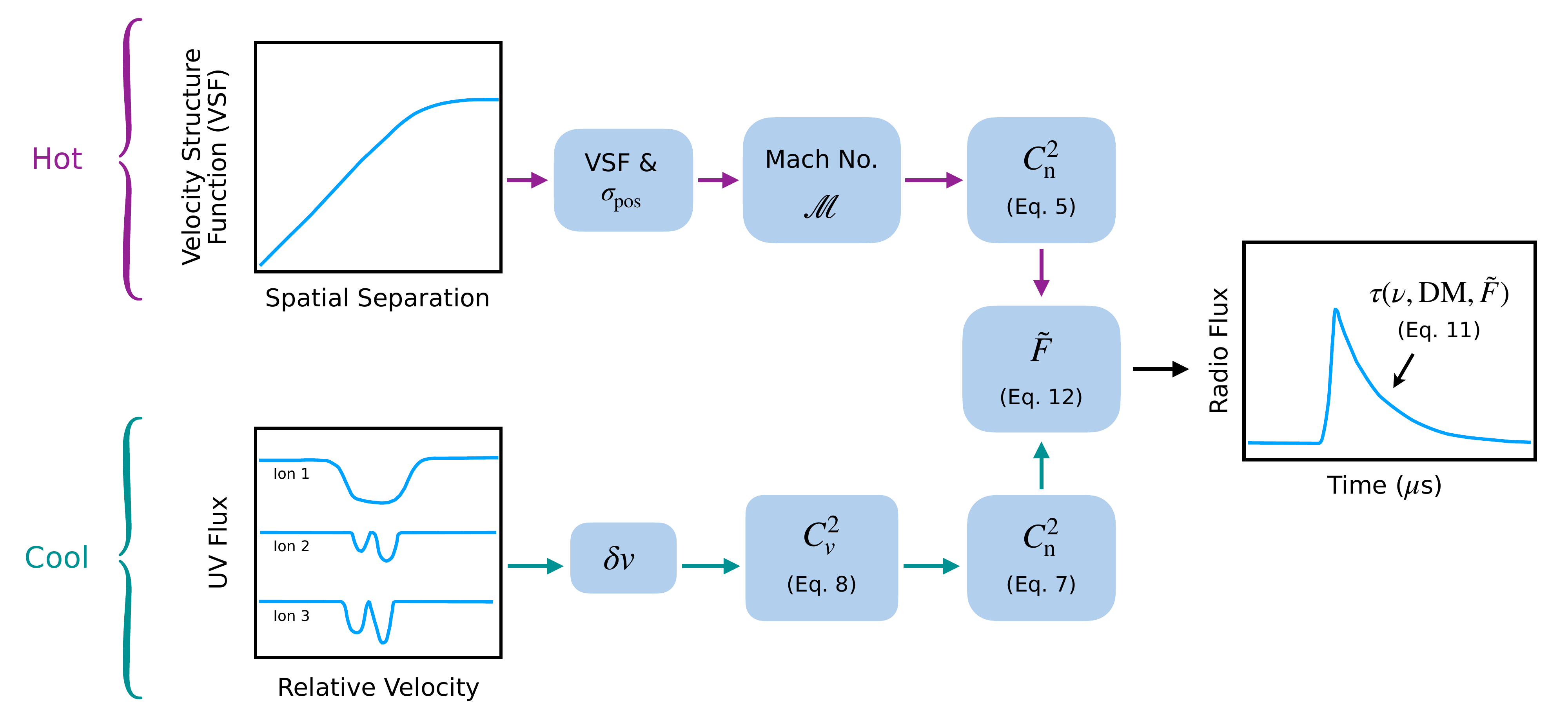}
    \caption{Schematic view of the methods used to derive density fluctuations from velocity fluctuation measurements in the CGM and subsequently predict levels of FRB scattering. All equations referenced are in Sections~\ref{sec:theory}-\ref{sec:obs}. See Section~\ref{sec:results} for further discussion.}
    \label{fig:methods_diagram}
\end{figure*}

{The formalisms laid out in Sections~\ref{sec:theory}-\ref{sec:obs} are used to convert velocity fluctuations into the density fluctuation amplitude $C_{\rm n}^2$ and subsequent strength of FRB scattering. Figure~\ref{fig:methods_diagram} illustrates our approach. For the hot CGM, Mach numbers inferred from VSFs {and projected velocity dispersion $\sigma_{\rm pos}$ }\citep{chen2023,chen2024} are converted to $C_{\rm n}^2$ for a range of possible densities, and assuming fiducial values for the outer scale noted further below (Equation~\ref{eq:Cn2-2}). For the cool CGM, nonthermal line widths indicate a typical velocity dispersion $\delta v \approx 22$ km/s at 1 kpc \citep{hchen2023}, which is used to estimate $C_v^2$ (Equation~\ref{eq:deltav-Cv2}), and subsequently $C_{\rm n}^2$ for a range of sound speeds and densities (Equation~\ref{eq:Cv-Cn2}). We then estimate $\tilde{F}$ from $C_{\rm n}^2$ for a range of possible dissipation scales $l_{\rm i}$ (Equation~\ref{eq:Ftilde-Cn2}). The resulting constraints on $C_{\rm n}^2$ and $\tilde{F}$ are presented in Section~\ref{sec:results1}, and are used to predict FRB scattering observables in Section~\ref{sec:results3}. These predictions are compared to observed FRBs in Section~\ref{sec:results2}. Prospects for constraining the dissipation scale of CGM turbulence are demonstrated in Sections~\ref{sec:results4}-\ref{sec:multipath}.} 

\subsection{Density Fluctuations in the Hot and Cool CGM}\label{sec:results1}

\begin{figure*}
    \centering
    \includegraphics[width=\textwidth]{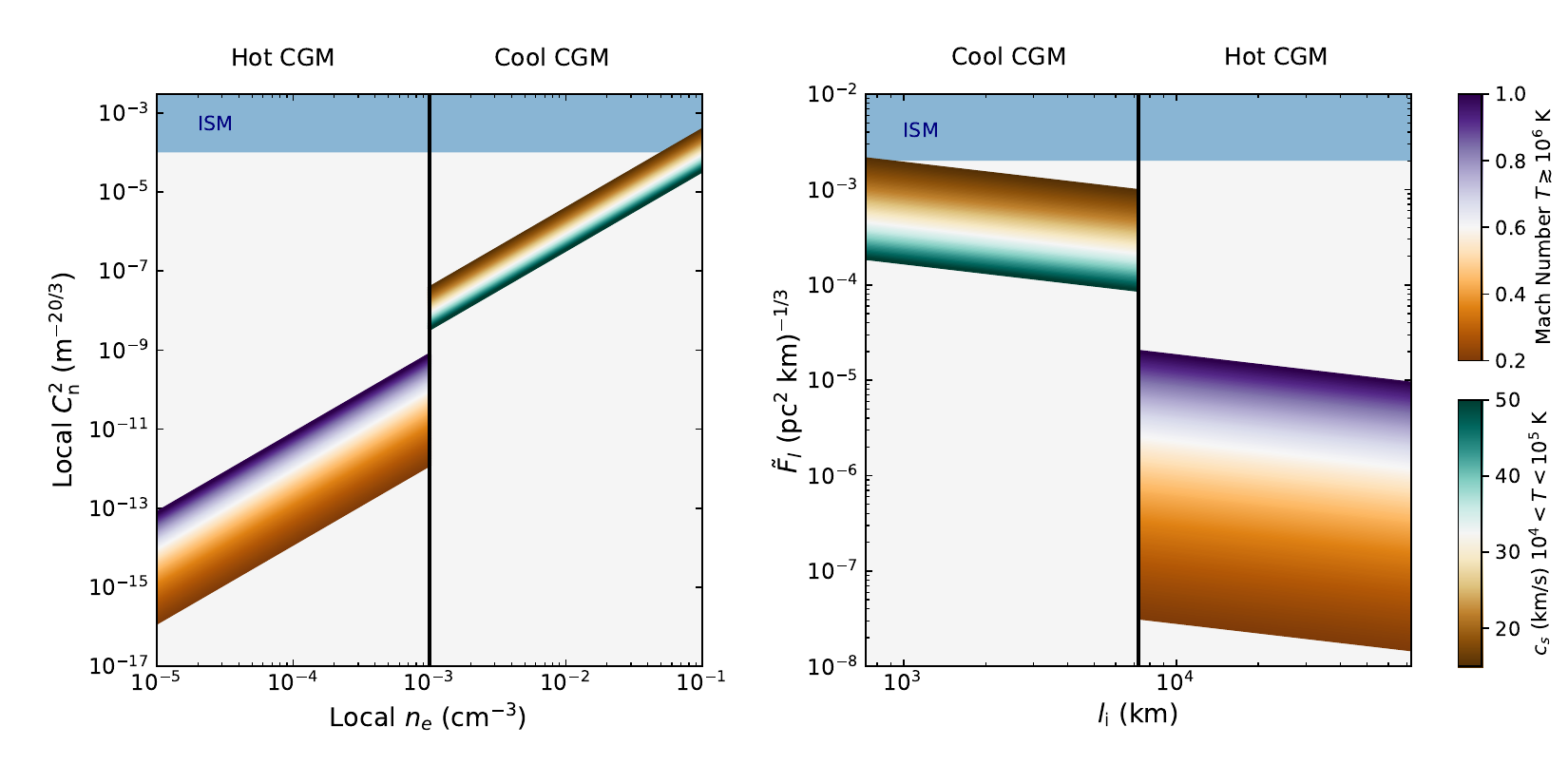}
    \caption{Density fluctuation spectral amplitude $C_{\rm n}^2$ vs. electron density (left) and the local density fluctuation parameter {$\tilde{F}_l \equiv f\times\tilde{F}$} {vs. inner scale} (right). For the hot CGM, $C_{\rm n}^2$ is evaluated using Equation~\ref{eq:Cn2-2} for a range of Mach numbers inferred from IFS observations of the hot CGM in emission. The outer (injection) scale for the hot CGM is taken to be $l_{\rm o} = 100$ kpc, and may be rescaled to different outer scales given $C_{\rm n}^2 \propto l_{\rm o}^{-2/3}$. For the cool CGM, $C_{\rm n}^2$ is evaluated using Equation~\ref{eq:Cv-Cn} for sound speeds corresponding to temperatures $10^4 \lesssim T \lesssim 10^5$ K. {The {local fluctuation parameter $\tilde{F}_l$} is evaluated from $C_{\rm n}^2$ for a range of inner scales equivalent to the ion inertial scale at the densities chosen for each phase (Equation~\ref{eq:Ftilde-Cn2}).} {The densities, $C_{\rm n}^2$, and $\tilde{F}_l$ shown for the cool gas represent values local to an individual cool gas structure, whereas the volume-averaged density and $\overline{C_{\rm n}^2}$ inferred from FRB measurements will be significantly smaller (and $\tilde{F}$ larger) for the entire halo due to the tiny volume-filling fraction of cool CGM gas. For the hot gas, $f \approx1$ and the local values shown here are good approximations of the volume-average.} The blue shaded regions show values typically inferred in the ionized ISM; in the ISM, $C_{\rm n}^2$ and $\tilde{F}$ can be even larger than the ranges shown.}
    \label{fig:Cn2_Ftilde} 
\end{figure*}

Figure~\ref{fig:Cn2_Ftilde} shows the density fluctuation spectral amplitude $C_{\rm n}^2$ vs. electron density for Mach numbers inferred in the hot CGM {and an outer scale $l_{\rm o} = 100$ kpc}. For a range of typical densities, $C_{\rm n}^2$ in the hot CGM is five to twelve orders of magnitude smaller than values typical of the ionized ISM probed by Galactic pulsars. Figure~\ref{fig:Cn2_Ftilde} also shows the equivalent range of $\tilde{F}$ vs. $l_{\rm i}$. Depending on the outer scale and Mach number, $\tilde{F}$ is between $\sim 10^{-8}$ (pc$^2$ km)$^{-1/3}$ to $\sim 10^{-4}$ (pc$^2$ km)$^{-1/3}$.

Constraints on $C_{\rm n}^2$ and {$\tilde{F}_l \equiv f\times\tilde{F}$} for the cool gas probed in absorption are also shown in Figure~\ref{fig:Cn2_Ftilde}. Here, $C_{\rm n}^2$ is estimated from $C_v^2$ for sound speeds  $15 \lesssim c_s \lesssim 50$ km/s (these are the sound speeds for $10^4 \lesssim T \lesssim 10^5$ K, the temperatures inferred from photoionization modeling; \citealt{hchen2023}), and for {local} densities between $10^{-3}$ and 0.1 cm$^{-3}$. While quasar absorption lines can constrain cloud densities $< 10^{-3}$ cm$^{-3}$, these smaller densities correspond to {a minority of absorbers and to} the largest cloud sizes ($>10$ kpc), for which modeling uncertainties are exacerbated. Due to the higher densities expected in the cool gas, $C_{\rm n}^2$ and $\tilde{F}_l$ reach values comparable to those routinely inferred in the ionized ISM. {It is important to note that the $C_{\rm n}^2$ values presented here for the cool gas are calculated from local electron densities in cool gas; the volume-averaged $C_{\rm n}^2$ ($\overline{C_{\rm n}^2}$) is $f \sim 10^{-3}-10^{-2}$ times smaller, and $\tilde{F}$ is $1/f$ times bigger than $\tilde{F}_l$. The volume-filling factor is correctly accounted for in Eq. \ref{eq:tau-Ftilde} for the mean scattering delay (see also Appendix \ref{app}).} 

\subsection{Scattering Predictions}\label{sec:results3}

\begin{figure}
    \centering
    \includegraphics[width=0.48\textwidth]{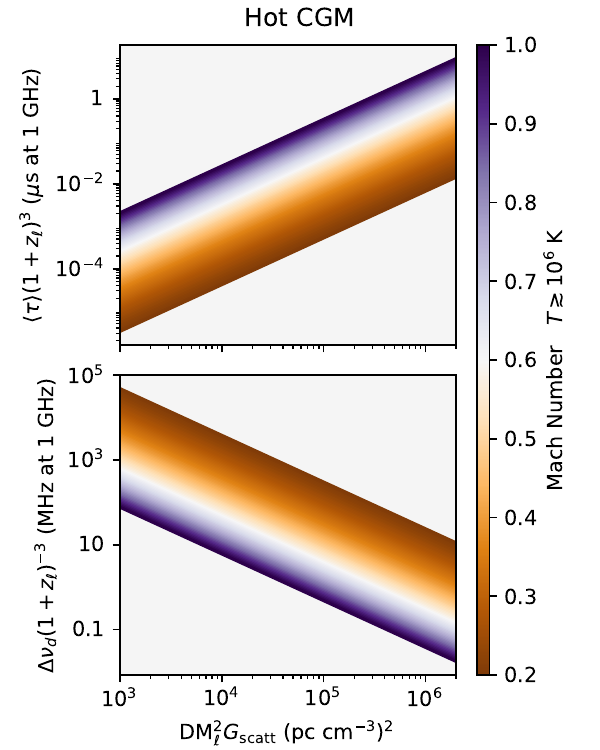}
    \caption{Mean pulse broadening time in the lens frame $\langle \tau \rangle (1+z_\ell)^3$ (top) and scintillation bandwidth in the lens frame $\Delta \nu_{\rm d}(1+z_\ell)^{-3}$ (bottom) vs. ${\rm DM}_\ell^2 G_{\rm scatt}$, a range of possible DMs and geometric configurations for the hot CGM of a halo intervening a LOS. Scattering parameters are estimated for Mach numbers typical of the hot CGM based on IFS velocity fluctuation measurements.}
    \label{fig:observables_hot}
\end{figure}

\begin{figure}
    \centering
    \includegraphics[width=0.48\textwidth]{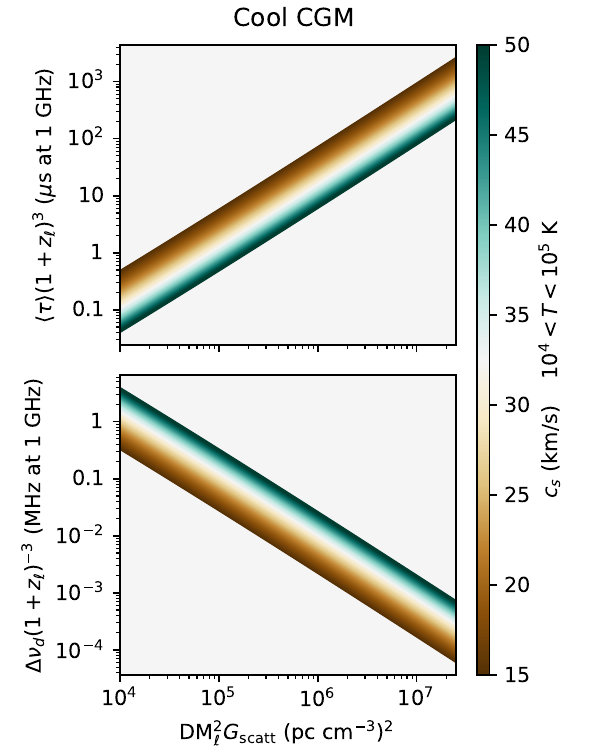}
    \caption{Same as Figure~\ref{fig:observables_hot} but for the cool CGM. Scattering parameters are estimated for sound speeds $c_s$ typical of the cool CGM using density fluctuations inferred from nonthermal broadening measurements of quasar absorption lines. {The range of ${\rm DM}_\ell^2 G_{\rm scatt}$ is based on densities and cloud sizes inferred for the quasar absorption line sample upon which our analysis is based \citep{hchen2023}; we assume $f\sim1$ and $\tilde{F}_l=\tilde{F}$ inside these clouds (see Section~\ref{sec:results3}).} }
    \label{fig:observables_cool}
\end{figure}

We use the $\tilde{F}$ derived from quasar constraints to estimate the {mean} pulse broadening time $\langle \tau \rangle$ and scintillation bandwidth $\Delta \nu_{\rm d}$ induced by the CGM of a single halo for a range of possible FRB LOS configurations. Figures~\ref{fig:observables_hot} and~\ref{fig:observables_cool} show $\langle \tau \rangle$ and $\Delta \nu_{\rm d}$ for the hot and cool CGM, respectively, in the rest frame of the scattering medium and for a wide range of ${\rm DM_\ell^2} G_{\rm scatt}$. {In the hot CGM, typical values of $\rm DM_\ell$ are $\sim100$ pc cm$^{-3}$  \citep{2019MNRAS.485..648P}, while $G_{\rm scatt}\sim 10^2-10^4$ for typical LOS geometries ($z\lesssim1$ for both FRBs and intervening galaxies, and path lengths $\sim0.1$ Mpc through a halo with order unity volume filling hot gas). For the cool CGM, photoionization modeling of quasar absorption lines yield densities $\sim10^{-3}-0.1$ cm$^{-3}$ for cloud sizes $\sim1$ pc to 1 kpc for the bulk of absorbers on which our $C_{\rm n}^2$ estimates are based \citep{hchen2023}. The corresponding DMs are $\sim0.1-1$ pc cm$^{-3}$ and $G_{\rm scatt} \sim 10^6-10^8$ (for the aforementioned cloud sizes and FRB redshifts $<1$). The resulting scattering predictions are similar to what would be inferred if we instead assumed a path length through the entire halo but a tiny volume filling factor for the cool gas (since $\tilde{F}$ would be $1/f$ times larger and $G_{\rm scatt}$ would be $1/L$ times smaller).}

{We find that $\langle \tau \rangle(1+z_\ell)^3$ is about $10^3$ times larger in the cool CGM than in the hot CGM. In the hot CGM, $\langle \tau \rangle(1+z_\ell)^3$ is less than $1$ $\mu$s at 1 GHz for a majority of plausible DMs and LOS configurations. In the cool CGM, $\langle \tau \rangle(1+z_\ell)^3$ spans $0.1$ $\mu$s to 1 ms at 1 GHz for the range of ${\rm DM}_\ell^2 G_{\rm scatt}$ considered, with the largest scattering delays produced by the smallest ($\sim10$ pc) clouds that have the greatest geometric leverage. Our inference from the quasar measurements suggests that turbulence in the hot CGM induces miniscule levels of pulse broadening, and that scattering from the cool CGM will only be detectable in the time domain when ${\rm DM}_\ell^2 G_{\rm scatt}$ is large.}

{These same results imply that scattering from the cool CGM may also be detectable in the frequency domain as diffractive scattering, as we find $\Delta \nu_{\rm d}(1+z_\ell)^{-3} \sim 0.01 - 1$ MHz at 1 GHz for about half of the range of ${\rm DM}_\ell^2 G_{\rm scatt}$ considered.} These scintillation bandwidths are readily measurable for typical observing configurations, with the main barrier to detection being the Galactic foreground, which can induce scintillation at comparable levels depending on the LOS, and the host galaxy ISM and/or immediate FRB environment, which may scatter the FRB sufficiently to quench CGM scintillation. Scintillation from the hot CGM could be detectable ($\Delta \nu_{\rm d}(1+z_\ell)^{-3} \sim 0.1 - 10$ MHz at 1 GHz) for large ${\rm DM_\ell}^2 G_{\rm scatt}$, but low Mach numbers and/or low ${\rm DM_\ell}^2 G_{\rm scatt}$ would correspond to $\Delta \nu_{\rm d}(1+z_\ell)^{-3} \gg 10$ MHz at 1 GHz, which may require the observing bandwidth of an ultra-wideband receiver. 

For typical source and lens separations, the $\langle \tau \rangle$ and $\Delta \nu_{\rm d}$ estimated for the cool CGM above are equivalent to angular broadening in the lens frame {$\theta_{\rm s} \sim (d_{\rm so}/d_{\rm sl})\theta_{\rm d} \sim 1 - 10$ $\mu$as} at 1 GHz. The hot CGM will produce even smaller $\theta_{\rm s}$. In the observer frame, these levels of angular broadening will likely be undetectable for most LOSs, due to both the Galactic foreground and typical radio imaging resolutions.

Scattering media near the source or observer have $G_{\rm scatt} \rightarrow 1$ \citep{cordes2022}. If the turbulence strengths inferred from quasar measurements are comparable to turbulence strengths in the Galactic CGM, then the expected amount of pulse broadening from the Galactic CGM is extremely small -- for expected DM contributions $\sim 10 - 100$ pc cm$^{-3}$ \citep{cook2023,dutta2024}, $\langle \tau \rangle$ is at the {microsecond} level or smaller, {below previously published upper limits \citep{ocker2021halo}.} Similar levels of scattering would also be expected for the host galaxy CGM.

\subsection{Comparison to Observed Scattering}\label{sec:results2}

The most stringent upper limits on scattering in the CGM are from FRBs with minimal scattering (based on observations with high time resolutions and/or large frequency bandwidths) and positively identified halos at known distances along their LOSs. About $15$ precisely localized FRBs in the literature have confirmed foreground halos with known redshifts \citep{flimflamdr1,vanLeeuwen23,faber24}. Only a few of these have measured scattering times or scintillation bandwidths comparable to the values estimated for the CGM above, {with the rest either having scattering significantly larger than expected for the CGM or lacking a scattering constraint altogether.}

Two of the most stringent upper limits on CGM scattering are from FRB20191108A (aka FRB191108) and FRB 20181112A (aka FRB181112). FRB191108 has a scintillation bandwidth of about 40 MHz at 1.37 GHz {(equivalent to 16 ns at 1 GHz for $\tau_{\rm d} \propto \nu^{-4.4}$)} and intersects M33 at an impact parameter of 18 kpc \citep{connor2020}. FRB181112 has a scattering time of $21\pm1$ $\mu$s at 1.3 GHz {(equivalent to 67 $\mu$s at 1 GHz}; \citealt{2020ApJ...891L..38C}) and intersects a foreground halo at $z_\ell = 0.36$ with an impact parameter of 29 kpc \citep{2019Sci...366..231P}. We subtract the Milky Way scattering contribution for these two FRBs by taking the smallest measured scattering time at 1 GHz of pulsars at similar absolute Galactic latitudes $|b|$, based on the pulsar scattering database in \cite{cordes2022}. For FRB181112 ($b = -48^\circ$), $\tau_{\rm min}^{\rm MW} \approx 1.9 $ ns at 1 GHz for pulsars near the same latitude, whereas for FRB191108 ($b = -30^\circ$), $\tau_{\rm min}^{\rm MW} \approx2.6$ ns at 1 GHz. The Galactic scattering contribution to these FRB LOSs will be three times larger {because when viewed from cosmological distances they behave like plane waves and are scattered through larger angles than Galactic sources, which are spherical waves \citep{cordes2016}.} The difference between the observed scattering and Milky Way scattering is then taken to be an upper limit on the total scattering from the CGM along the FRB LOS; assuming a small Milky Way contribution thus yields a larger and more conservative upper limit on the scattering contribution of the intervening CGM.

\begin{figure}
    \centering
    \includegraphics[width=0.48\textwidth]{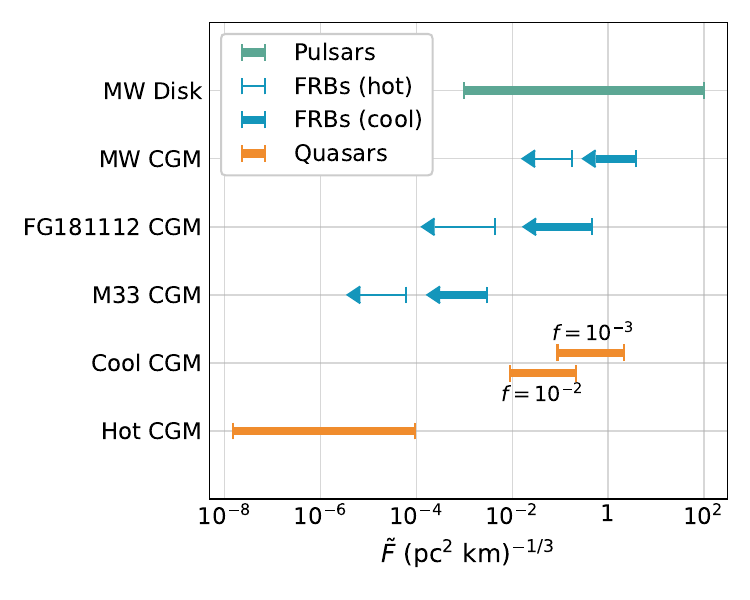}
    \caption{The density fluctuation parameter $\tilde{F}$ inferred for the CGM and Milky Way disk from pulsar (green), FRB (blue), and quasar (orange) observations. The range of values shown for the Milky Way disk is based on pulsar scattering measurements, while the upper limit on $\tilde{F}$ in the Milky Way halo is based on two repeating FRBs for which a scattering budget can be constructed \citep{ocker2021halo}. Two other upper limits on $\tilde{F}$ from FRBs are shown, {the foreground galaxy CGM intersected by FRB181112 and the M33 CGM intersected by FRB191108.} The FRB constraints are separated into the $\tilde{F}$ that would be obtained if all of the scattering is attributed to cool gas (thick lines) or to hot gas (thin lines), based on the different expected DMs, {and after subtracting the minimum scattering contribution of the Milky Way} (see Section~\ref{sec:results2} {and Equation~\ref{eq:tau-Ftilde}}). {FRB and pulsar constraints account for the conversion between the $1/e$ delay measured from pulse shapes and the mean scattering time; for FRB scattering in the CGM, the assumed conversion factor is $A_\tau \approx 1/6$, whereas for pulsars we assume $A_\tau \approx 1$ because their larger scattering leads to smaller diffractive scales and hence larger $A_\tau$.} The range of $\tilde{F}$ shown for the cool and hot CGM traced by quasars is based on Figure~\ref{fig:Cn2_Ftilde} above, and assumes an inner scale equal to the ion inertial scale. {For the cool gas, local estimates $\tilde{F}_l$ are converted to global values applicable to an entire halo, $\tilde{F} = \tilde{F}_l/f$, for two possible filling factors ($f=10^{-3}$ and $f=10^{-2}$).}}
    \label{fig:Ftilde_summary}
\end{figure}

\cite{ocker2021halo} used these scattering measurements to derive upper limits on $\tilde{F}$ in the CGMs of the foreground galaxies, in addition to deriving $\tilde{F}$ in the Milky Way disk and CGM using Galactic pulsars and two repeating FRBs with precise scattering budgets. Figure~\ref{fig:Ftilde_summary} shows $\tilde{F}$ inferred for the Galactic CGM and disk based on these FRB and pulsar measurements, along with the $\tilde{F}$ we infer for the hot and cool CGM based on quasars. {We convert the local $\tilde{F}_l$ estimated for the cool CGM from quasar absorption lines into volume-averaged $\tilde{F} = \tilde{F}_l/f$ for two possible filling factors of the cool gas, $f=10^{-3}$ and $f=10^{-2}$. In this way, all $\tilde{F}$ values shown in Figure~\ref{fig:Ftilde_summary} represent global quantities that can be directly compared.}

\cite{ocker2021halo} estimated $\tilde{F}$ in the CGM by assuming the entire column density of hot halo gas contributed to scattering, and used a modified Navarro-Frenk-White (mNFW) profile \citep{2019MNRAS.485..648P} with a cutoff at twice the virial radius to estimate the corresponding DM. Here, we instead separate the CGM constraints from \cite{ocker2021halo} into the $\tilde{F}$ that would be inferred if scattering were attributed to either the hot CGM or the cool CGM. For the hot CGM, we use the same mNFW profile and assumptions as in \cite{ocker2021halo}, which yield DM contributions of about 90 pc cm$^{-3}$ for FRB191108 and 135 pc cm$^{-3}$ for FRB181112. 

For the cool gas, column densities in neutral hydrogen are well constrained observationally from quasar absorption lines, and yield typical hydrogen column densities $N_{\rm H} \sim 10^{18} - 10^{20}$ cm$^{-2}$ for neutral fractions $\sim 10^{-2} - 10^{-3}$ \citep{werk2014}. We use the mean $N_{\rm H}$ of the COS-Haloes sample, $10^{19.6}$ cm$^{-2}$ \citep{werk2014}, equivalent to a DM of 13 pc cm$^{-3}$. For the Galactic cool CGM, we use a mean DM of 20 pc cm$^{-3}$ inferred from HI measurements of high-velocity clouds \citep{2019MNRAS.485..648P}. While there is considerable uncertainty in the DM contribution of the CGM, these fiducial values sufficiently illustrate how the range of $\tilde{F}$ compares between the different tracers shown in Figure~\ref{fig:Ftilde_summary}. {Similar to the FRB upper limits in the hot gas, we use the entire path length through the halo to estimate $G_{\rm scatt}$, so that all $\tilde{F}$ values shown in Figure~\ref{fig:Ftilde_summary} are comparable volume-averaged quantities.}

{The upper limits on $\tilde{F}$ in the Milky Way CGM and foreground CGM of FRB181112 are consistent with the range of values inferred from quasar measurements for the cool CGM, and are too large to be attributable to the hot CGM. However, for FRB191108 the upper limit on $\tilde{F}$ derived for cool gas is smaller than the range of values we infer from quasars for volume filling factors $f\sim10^{-3}-10^{-2}$; an unreasonably large filling factor ($f\sim0.1$) would be required to make the quasar-based prediction consistent with the FRB-based upper limit. On the other hand, the upper limit estimated for FRB191108 is consistent with the range of $\tilde{F}$ estimated from quasars for the hot gas.}

{The scattering of FRB181112 was explored in detail in \cite{2019Sci...366..231P}, who used the measured $1/e$ delay to infer $r_{\rm diff}$, and subsequently SM and $\overline{C_{\rm n}^2}$ (Equation~\ref{eq:rdiff}). Applying their methodology to the more up-to-date measurement $\tau_{\rm d} = 21$ $\mu$s \citep{2020ApJ...891L..38C} yields $\rm SM \sim 9.1\times10^{-6}$ kpc m$^{-20/3}$ and $\overline{C_{\rm n}^2} \sim 10^{-7}\times(L/100\ {\rm kpc})^{-1}$ m$^{-20/3}$ (slightly larger than the values quoted in \citealt{2019Sci...366..231P}). Note this is the average $\overline{C_{\rm n}^2}$ along a LOS of length $L$ through the halo. This value of $\overline{C_{\rm n}^2}$ is significantly larger than that expected for the hot gas based on quasars, but comparable to the largest values expected in the cool gas. For a cool volume filling fraction $\sim10^{-3}$ and a local density $n_e\sim0.1$ cm$^{-3}$, $\overline{C_{\rm n}^2}\sim10^{-7}$ m$^{-20/3}$ (Equation~\ref{eq:Cv-Cn2}), comparable to the value estimated from $\tau_{\rm d}$. Our inference from quasar observations thus supports the interpretation that if the scattering of this FRB occurs in the CGM, then it is likely dominated by the cool gas.}

\subsection{The Dissipation Scale of CGM Turbulence}\label{sec:results4}

Consistency between the $\tilde{F}$ inferred from FRB scattering and quasar-based velocity fluctuation measurements does not necessarily translate into positive proof that CGM density fluctuations extend to $\lesssim$ au scales (after all, we have assumed a single uniform cascade of density fluctuations across sub-au to $>$kpc scales, when in fact there may be multiple injection and dissipation scales, which can also alter the shape of the density fluctuation spectrum and introduce deviations from Kolmogorov turbulence). However, since $\tilde{F}$ depends on the inner scale $l_{\rm i}$, in principle one can combine FRB-based constraints on $\tilde{F}$ (which are based on observables and make no assumption about $l_{\rm i}$) with quasar-inferred values of $\tilde{F}$ (which necessarily cover a range of possible $l_{\rm i}$), to yield either a lower limit or, with complete information, a constraining inference of the inner scale.

\begin{figure*}
    \centering
    \includegraphics[width=0.9\textwidth]{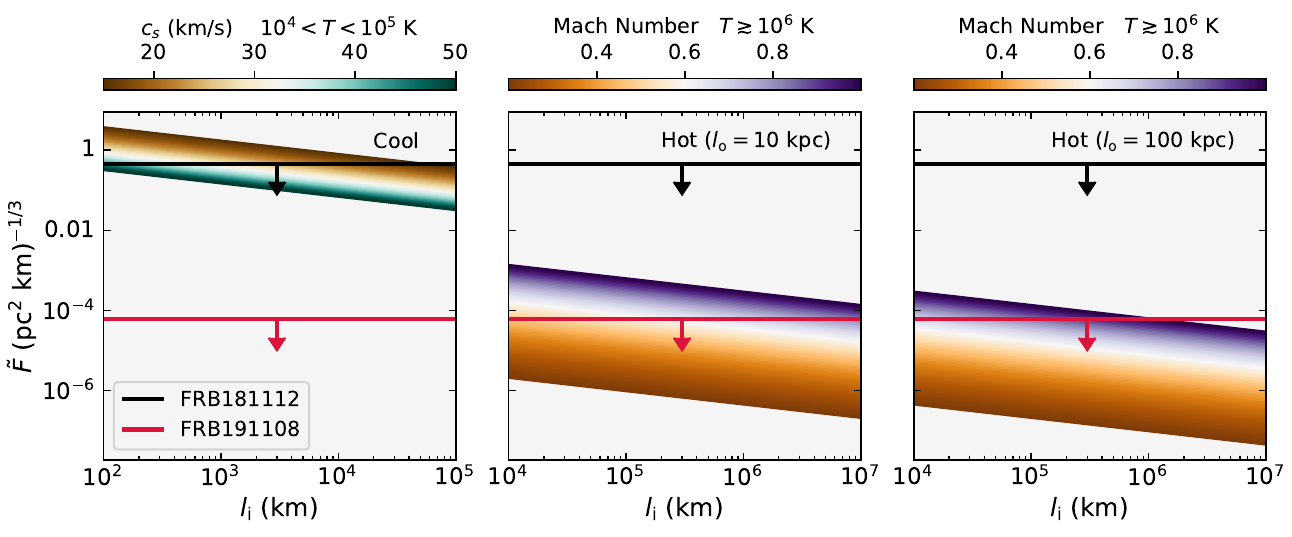}
    \caption{Density fluctuation parameter $\tilde{F}$ vs. the dissipation scale, for the cool (left) and hot (middle, right) CGM as inferred from quasar observations. {Upper limits on $\tilde{F}$ from FRB181112 and FRB191108 are shown by the black and red horizontal lines, respectively, separated into constraints for the cool (FRB181112) and hot (FRB191108) phases based on their different DM contributions,} {and after subtracting the minimum Milky Way scattering contribution} (see Section~\ref{sec:results2}). {Values of $\tilde{F}$ shown for the cool gas assume $f\sim10^{-3}$; a smaller filling factor would increase $\tilde{F}$ by a factor $1/f$.} }
    \label{fig:inner_scale_limits}
\end{figure*}

Figure~\ref{fig:inner_scale_limits} shows $\tilde{F}$ vs. {a wide range of possible values for the} inner scale, based on the $\overline{C_{\rm n}^2}$ inferred for the hot and cool CGM in Figure~\ref{fig:Cn2_Ftilde}. These quasar-based constraints on $\tilde{F}$ are compared to {FRB181112 and FRB191108, which give upper limits on $\tilde{F}$ irrespective of the inner scale}. Taking this comparison at face value, the inner scale of the coldest ($\sim 10^4$ K) gas would need to be {$\gtrsim 750$ km for a sound speed $c_s\approx30$ km/s,} in order for the quasar and {FRB181112} constraints to be consistent; this constraint falls between the electron inertial length and ion gyroradius typical of the cool CGM (Table~\ref{tab:microscales}). For the hot gas, the {FRB191108} upper limit has more constraining power if the outer scale is small ($l_{\rm o} = 10$ kpc), and is consistent with an inner scale $\gtrsim 10^4$ km for Mach numbers $\lesssim 0.5$. If the outer scale is large ($l_{\rm o} \sim 100$ kpc) and $\mathcal{M} \approx 1$, then $l_{\rm i} \gtrsim 10^6$ km, {much greater than the ion inertial length}.

{These lower limits are purely demonstrative and do not account for potentially large uncertainties in the DM contribution of the CGM, which dominates the total uncertainty in the constraint on $l_{\rm i}$ since $l_{\rm i} \propto \rm DM^{-6}$. A full error accounting should also include formal uncertainty errors on the scattering measurement, uncertainty in the Milky Way scattering subtraction, {and uncertainty in the cool gas volume filling factor}. For the purposes of this initial comparison, we note that reducing the Milky Way scattering contribution to an even smaller value, e.g. 1 ns, would yield a lower limit on $l_{\rm i}$ in the cool gas that is more comparable to the electron inertial length, but still compatible with dissipation at plasma microscales.}

In principle, this comparison can be highly informative when the sound speed or Mach number is either known or more tightly bound, and when the FRB and quasar constraints are placed on the same system. Hydrogen column densities inferred from quasar absorption lines could also be used to place tighter constraints on the DM contribution of the CGM -- albeit along a different line of sight -- mitigating uncertainty in the FRB DM budget. If the CGM in question has detectable emission, then IFS measurements could also be used to infer the turbulence spectral index, and the subsequent analysis can be modified accordingly (rather than assuming a Kolmogorov spectral index as we have done here), though of course this still involves extrapolation over many orders of magnitude, along which the spectral index could change. 

\subsection{Conditions for Multipath Propagation}\label{sec:multipath}

The results above all assume that the conditions for multipath propagation are satisfied in the CGM, i.e., that the CGM does produce some level of radio pulse broadening or scintillation. If, however, such scattering from the CGM can be firmly ruled out, then the inner scale of the density fluctuation spectrum must be greater than the multipath scale (Equation~\ref{eq:rmp}).
{CGM scattering could be ruled out by localization of scattering screens to the Milky Way and/or host galaxy ISMs, which requires some combination of pulse broadening, scintillation bandwidth, and/or angular broadening measurements. Detection of scintillation over months-long baselines (e.g., for a repeating FRB), or combining an angular broadening measurement with a scintillation bandwidth, can also constrain the scattering screen location, in some cases revealing it to be within the Milky Way ISM \citep{ocker2021halo,wu2024}. When pulse broadening or scintillation is detected on distinct frequency scales, such that the observed scattering cannot be explained by a single screen, the scattering screens can typically be localized to within the Milky Way and host galaxy \citep{2015Natur.528..523M,ocker2022,sammons2022,nimmo2025}. Such an analysis on an FRB with an intervening CGM would be the basis for determining whether the scattering is consistent or inconsistent with arising in that CGM.}

\begin{figure}
    \centering
    \includegraphics[width=0.48\textwidth]{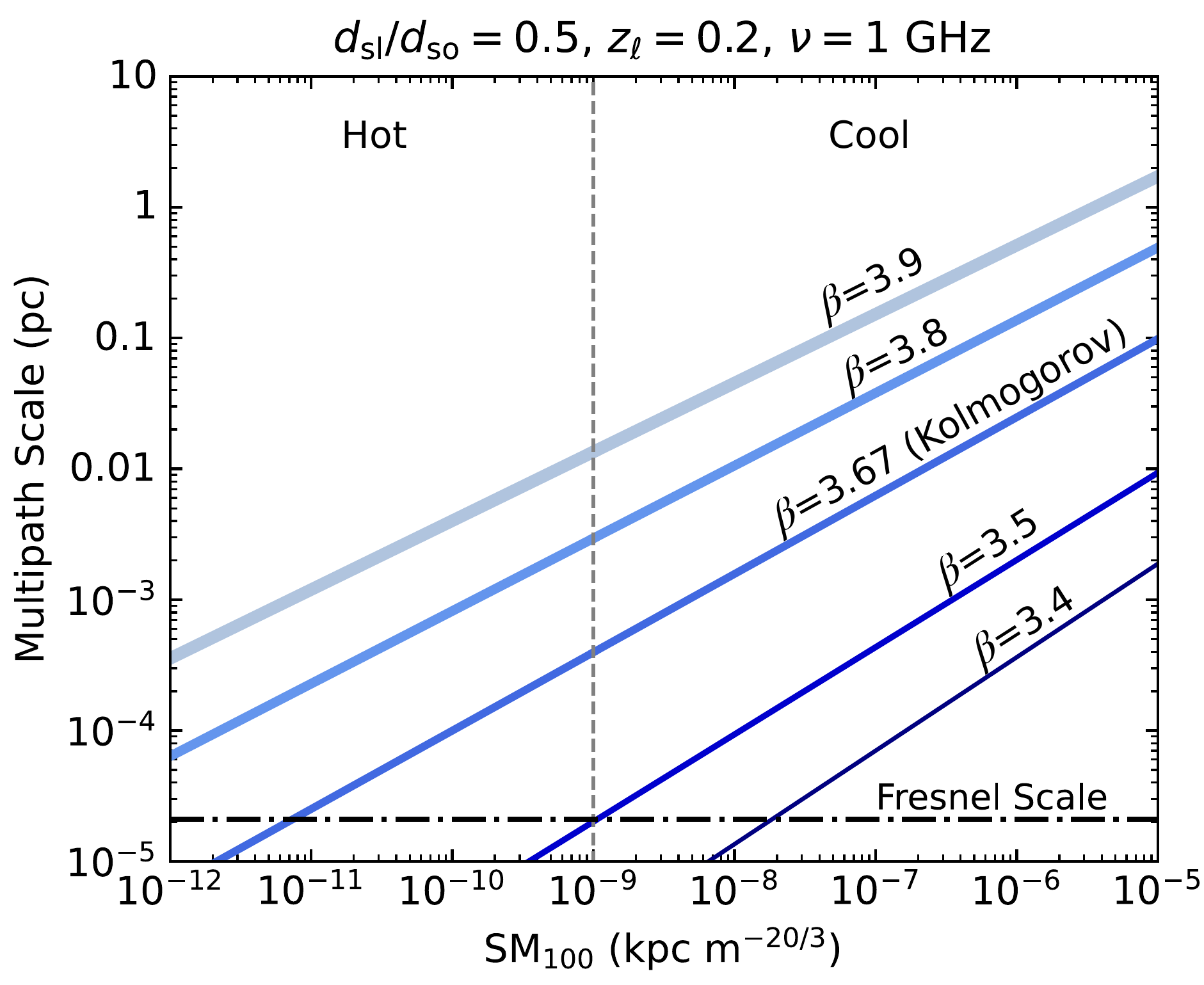}
    \caption{The multipath scale $r_{\rm mp}$ (equivalent to the scattering cone diameter at the screen) vs. {${\rm SM}_{100} \equiv \overline{C_{\rm n}^2} \times 100$ kpc}, for the range of $C_{\rm n}^2$ found in the CGM (Figure~\ref{fig:Cn2_Ftilde}) {and assuming $f=1$ in the hot gas and $f=10^{-3}$ in the cool gas}. 
    Here $r_{\rm mp}$ is evaluated as $r_{\rm mp} = 2d_{\rm lo}\theta_{\rm d} = (d_{\rm lo}d_{\rm sl}/d_{\rm so}) \lambda/\pi r_{\rm diff}$, where $r_{\rm diff}$ is defined as in Equation~\ref{eq:rdiff}, for fiducial values $d_{\rm sl}/d_{\rm so} = 0.5$, $\lambda = 0.3$ m, $z_\ell = 0.2$, and a range of $\beta$. The approximate division between ${\rm SM_{100}}$ characteristic of hot and cool gas is indicated by the grey dashed line. 
    {When $r_{\rm mp} < r_F \sim 2\times10^{-5}$ pc (black dash-dotted line), the strong scattering regime no longer applies.}} 
    \label{fig:multipath_scale}
\end{figure}

Figure~\ref{fig:multipath_scale} shows $r_{\rm mp}$ vs. {${\rm SM}_{100} \equiv \overline{C_{\rm n}^2} \times 100$ kpc}, the scattering measure for the range of $C_{\rm n}^2$ extrapolated from quasar measurements and a fiducial path length of 100 kpc. {Local values $C_{\rm n}^2$ estimated from quasars are converted to $\overline{C_{\rm n}^2} = fC_{\rm n}^2$ assuming $f=1$ in the hot gas and $f=10^{-3}$ in the cool gas.} We assume the same path length for both gas phases, noting that $\rm SM_{100}$ can be easily rescaled; e.g., for a path length of 10 kpc $\rm SM_{100}$ would be 10 times smaller than the values shown in Figure~\ref{fig:multipath_scale}. {Only multipath scales greater than the Fresnel scale are relevant, because when $r_{\rm mp} < r_F$ the strong scattering regime (and the scattering formulas used in this work) no longer apply.} The multipath scale is significantly affected by the choice of $\beta$: For a given $\rm SM_{100}$, $r_{\rm mp}$ differs by {three orders of magnitude between $\beta = 3.4$ and $\beta = 3.9$}. By contrast, modifying $\lambda$ and $z_\ell$ within their typical ranges ($0.1 \lesssim \lambda \lesssim 0.8$ m, $z_\ell < 1$) changes $r_{\rm mp}$ by less than an order of magnitude from the values shown in Figure~\ref{fig:multipath_scale}. For a fixed path length of 100 kpc, $r_{\rm mp}$ in the hot CGM ranges from {a few au to $\sim 0.01$ pc}, depending on $\beta$. {If $\beta$ is small ($\lesssim 3.4$), $r_{\rm mp}$ is comparable to the Fresnel scale and no strong diffractive scattering will occur regardless of the inner scale -- in this scenario, a scattering non-detection would not yield any information on the inner scale.} For $\beta = 3.67 \approx 11/3$ (Kolmogorov), $r_{\rm mp}$ can be larger than the Fresnel scale, and a robust non-detection of scattering in this regime would suggest an inner scale much larger than plasma microscales, {which could imply the presence of a damping process, e.g., {cooling or} ion-neutral damping, that terminates the cascade above kinetic scales \citep{lithwick2001}}. For {$\rm SM_{100} \gtrsim 10^{-9}$ kpc m$^{-20/3}$}, roughly characteristic of the cool CGM, $\beta \approx 11/3$ yields {$r_{\rm mp} \sim 10^{-3} - 0.1$ pc}, comparable to the smallest spatial scales inferred for cool CGM clouds from quasar absorption lines (although these cloud sizes may be overestimated due to the limited velocity resolution of spectrographs; \citealt{rudie2019,hchen2023}). This result suggests that if the cool CGM produces scattering, then cool CGM cloudlets must sustain internal turbulence at scales much smaller than their sizes.

\section{Discussion}\label{sec:discuss}

{We have used quasar observations of CGM turbulence to infer the strength of density fluctuations in both the hot and cool CGM, and subsequent levels of FRB scattering. This analysis implicitly requires several key assumptions, which are discussed in detail below. Physical implications of our results, including in comparison to previous predictions of FRB scattering in the CGM, are also discussed. Conclusions are summarized in Section~\ref{sec:conc}.}

\subsection{Modifying Our Assumptions}

While we have assumed uniform, isotropic Kolmogorov turbulence for most of our analysis, there is evidence that CGM turbulence can deviate from a Kolmogorov spectrum, and that the turbulence spectrum may include local enhancements due to shocks and/or local energy sources (e.g., inflows and outflows). Turbulence spectra significantly shallower or steeper than Kolmogorov would change our scattering predictions. Keeping $\delta n_e$ fixed at the outer scale (to, e.g., a value set by the Mach number), $3 < \beta < 11/3$ will yield $C_{\rm n}^2$ larger than the values we estimate for $\beta = 11/3$, thus leading to larger scattering delays for a given halo and LOS geometry {(provided that the multipath scale remains larger than the Fresnel and diffractive scales)}. Conversely, $11/3 < \beta < 4$ will yield $C_{\rm n}^2$ and $\langle \tau \rangle$ smaller than what we find for $\beta = 11/3$. For $\beta > 4$ refractive effects will dominate over a larger range of spatial scales and the diffractive scattering considered here may become irrelevant {because density fluctuations would become too weak at the length scales relevant to diffraction} \citep{goodman85,cordes1986}. If $\beta$ is well determined from velocity fluctuation measurements, then its empirical value should be used to estimate $C_{\rm n}^2$, although this still involves extrapolation over many orders of magnitude.

{All of our scattering predictions (pulse broadening times and scintillation bandwidths) assumed by necessity that density fluctuations dissipate at scales much smaller than the Fresnel scale ($\sim 10^{13}$ cm). Isothermal turbulence in the cool gas may be damped by cooling below the cooling length scale \citep{lithwick2001}, although this effect has not yet been shown in modern, large-scale numerical simulations \citep[e.g.][]{beattie2025}. In the absence of either damping by cooling or neutrals, \cite{lithwick2001} argue that density fluctuations will cut off below the transverse proton diffusion scale, which is $l_{\rm pd}^{\perp} \sim 10^{10}$ cm in the cool gas (comparable to $r_{\rm diff}$) and $l_{\rm pd}^{\perp} \sim 10^{17}$ cm in the hot gas (larger than $r_{\rm mp}$). 
It is possible that density fluctuations in the hot gas are damped well above the spatial scales relevant to radio wave scattering. However, observational evidence suggests that density fluctuations can survive below $l_{\rm pd}^{\perp}$, down to kinetic scales \citep{schekochihin2009,2019NatAs...3..154L}.}

{Regardless of the exact value of the inner scale, scattering observations provide a means of distinguishing the role of different damping/dissipation mechanisms in magnetic turbulence, the nature of which is poorly understood in the CGM. If the inner scale is $\gtrsim r_{\rm mp}$, then no multipath propagation (be it diffractive or refractive) will be observed. An inner scale between $r_F$ and $r_{\rm mp}$ could yield refractive multiple imaging (but not diffractive scattering). If the inner scale is comparable to or greater than $r_{\rm diff}$, then the scattered image will be close to Gaussian and the shape of the scattered pulse will approach a one-sided exponential. An inner scale much smaller than $r_{\rm diff}$, as assumed throughout this paper, will yield a scattered pulse that deviates from a one-sided exponential. This latter case has been observed in the Galactic warm ionized ISM \citep{rickett2009,geiger2024}, and estimates of $r_{\rm diff}$ and kinetic scales may support a similar scenario in the CGM.}

In addition to the assumptions discussed above, we have adopted a range of sound speeds and Mach numbers based on the range of CGM systems with observed turbulence, none of which overlap with known FRBs. Targeted surveys of FRB-quasar pairs intersecting foreground galaxies, or FRBs intersecting foreground CGMs detected in emission, may directly yield estimates of the Mach number, $\beta$, and even $l_{\rm o}$ in those individual systems, enabling a more direct comparison between FRB and quasar observations. Such a comparison may already be achievable in M31, which covers a large enough area of the sky that dozens of quasars and FRBs intersect it \citep{2020ApJ...900....9L,connor2022_halos}. 

\subsection{Dissecting the Cloudlet Model: Volume Filling Fraction and Outer Scale of the Cold CGM}

Our formalism for relating scattering observables to density fluctuations is based on an ionized cloudlet model in which the CGM is composed of cloudlets that sustain internal turbulence, while allowing for fluctuations in the mean density of different cloudlets. Our analysis suggests that cooler ($10^4 - 10^5$ K) cloudlets are the most likely to produce observable FRB scattering, as suggested in previous studies. One may then ask what additional properties of the cool CGM may be constrained by our model, other than the strength of density fluctuations and their dissipation scale. The size of individual cloudlets is not explicitly factored into the model, but it is implicitly related to the model parameters $\zeta$, $f$, and $l_{\rm o}$. If the CGM is composed of a single large cloud with uniform mean density, then $\zeta = \langle n_{ec}^2\rangle/\langle n_{ec}\rangle^2$ and the volume filling factor $f$ will be unity, with the path length through the scattering screen $L$ comparable to the cloud size. If, on the other hand, the CGM is significantly ``clumpy,'' with a range of mean cloudlet densities along the LOS, then $\zeta > 1$ (this parameter is identical to a clumping factor; e.g. see Eq. 18 in \citealt{dutta2024}). If cool gas is additionally concentrated into numerous tiny cloudlets spread uniformly in the CGM with a small volume fraction $f \ll 1$, $L$ is comparable to the virial radius. A small volume filling fraction with a large area covering fraction of cool gas is implied by quasar absorption studies (see \citealt{hummels2024,dutta2024,singh2024} for a range of possible models). Regardless of the exact scenario, the cloudlet size may also be related to the outer scale of the density fluctuations within cloudlets. These parameters are all lumped together in the composite $\tilde{F} = \zeta \epsilon^2/f (l_{\rm o}^2 l_{\rm i})^{1/3}$ {and $\tilde{F}_l \equiv f\times\tilde{F}$}, which we infer both from quasar observations of turbulence and upper limits on FRB scattering in the CGM.  

While dissecting $\tilde{F}$ is difficult, involving five degenerate parameters, it is not altogether impossible. In principle, $\epsilon^2$ and $l_{\rm o}$ can both be inferred by leveraging quasar-based constraints on column densities and cloud sizes, and $l_{\rm i}$ can be constrained through the methods laid out in this paper. {Given some prior on $\epsilon^2$, $l_{\rm o}$, and $l_{\rm i}$, along with an inference of $\tilde{F}_l \equiv f\times\tilde{F}$, one can then constrain the likely range of $\zeta$, which will be close to unity for a highly uniform medium or much greater than unity for a very clumpy medium. As an illustration, if we assume $n_e \sim10^{-2}$ cm$^{-3}$ and $l_{\rm o} \sim 10$ pc, we have $C_{\rm n}^2 \sim 10^{-6}$ m$^{-20/3}$ (Figure~\ref{fig:Cn2_Ftilde}) and $\epsilon^2 \sim 10^{-4}$ (Equation~\ref{eq:dne-Cn2}). For $l_{\rm i} \sim 2000$ km (the ion inertial scale for $n_e \sim 10^{-2}$ cm$^{-3}$), and a total constraint on $\tilde{F}_l \sim 5\times10^{-4}$ (pc$^2$ km)$^{-1/3}$ (Figure~\ref{fig:Cn2_Ftilde}), we would then infer $\zeta \sim 300$. Conversely, assuming a perhaps unphysically large outer scale $l_{\rm o} \sim 100$ kpc would imply $\zeta \approx 1$, implying an unrealistically smooth medium for any plausible range of cold gas overdensities. While these are purely demonstrative calculations, they illustrate that our empirically motivated constraints on $\tilde{F}_l$ would imply insensibly little clumpiness of high-density gas if an outer scale comparable to the halo radius is assumed. The line of reasoning followed above offers one possible path to interpreting quasar and/or FRB-based constraints on $\tilde{F}_l$ (or $\tilde{F}$).}

\subsection{Comparison with VP19 and the Misty CGM}

Applying this ionized cloudlet model to quasar-based inferences of turbulence yields scattering predictions that, {depending on sightline geometry, are up to $10^3$ times} smaller than the fiducial values predicted by \cite{2019MNRAS.483..971V} (hereafter VP19), who used a simple analytic prescription for the \cite{2018MNRAS.473.5407M} misty CGM, in which the CGM is a volume of cool cloudlets with a volume filling factor $f \sim 10^{-4}$, internal density $n_{e} \sim 10^{-3}$ cm$^{-3}$, and cloudlet radius $r_c \sim 1$ pc. In VP19, scattering arises from the random distribution of cloudlets along a given LOS, with the superposition of intersected cloudlets yielding phase perturbations that mimic a Kolmogorov spectrum with an outer scale comparable to the cloudlet size. The smallest scale cutoff of the phase perturbations (i.e., the inner scale) is not included in their model, because they focus on predicting the diffractive scattering time, which depends on $r_{\rm diff}$ {and is independent of $l_{\rm i}$}. While their model sources density fluctuations not from turbulence, but from the number and mean density of intersected cloudlets, making a direct comparison between our results difficult (and perhaps unphysical), we can nonetheless investigate what modifications to their model would yield scattering predictions more similar to our own.

In the VP19 model, $\tau_{\rm d} \propto r_c^{-2} \nu^{-4.4} {\rm DM}^{2.4} f_a^{1.2} (d_{\rm sl}d_{\rm lo}/d_{\rm so})$, where $r_c$ is the cloud size and $f_a \sim fb/r_c$ is the number of intersected clouds at an impact parameter $b$, for a volume-filling factor $f$. Their fiducial values $r_c\sim1$ pc, $\nu \sim 1$ GHz, $\rm DM \sim 0.03$ pc cm$^{-3}$, $f_a\sim10$, and $(d_{\rm sl}d_{\rm lo}/d_{\rm so})\sim 1~{\rm Gpc}$ yields $\tau_{\rm d}\sim0.4$ ms. 
Since decreasing $\tau_{\rm d}$ requires increasing $r_c$ and/or decreasing $f_a$, the most natural way to produce a smaller scattering prediction is to thus increase $r_c$. A modest increase from $r_c\sim1$ pc to $r_c\sim 10$ pc is sufficient to make VP19's predictions broadly consistent with our own (see their Equation 19), {if we account for the corresponding increase in DM}.

An important limitation of VP19 is the assumption of a universal cloud size. A key motivation for considering a misty CGM is the large area covering factor $f_A \sim \mathcal{O}(1)$ of cold gas, despite its small volume filling fraction $f\sim10^{-3}$ \citep{2018MNRAS.473.5407M}. Since $f_{\rm A} \sim (R/r_c) f$, for an $R \sim 100$ kpc halo to have order unity $f_A$ we require $r_c < 100$ pc. \citet{2018MNRAS.473.5407M} argued that the cooling length $l_{\rm s} \sim {\rm min}(c_s t_{\rm cool})$, the scale on which cooling clouds establish sonic contact with their surroundings, sets a characteristic cloud length scale. Since $l_{\rm s} \propto n_e^{-1}$, this corresponds to a fixed column $N_{\rm e} \sim 10^{17}~{\rm cm}^{-2}$, or $l_{\rm s} \sim 30~{\rm pc} \, (n_e/10^{-3})^{-1} {\rm cm}^{-3}$. This characteristic scale is much larger than $r_c\sim 1$ pc clouds frequently invoked in the literature, which is actually the value of $r_c$ for $l_{\rm s}$ in higher redshift, denser halos\footnote{\citet{2018MNRAS.473.5407M} focused on such systems, as they found that-- by comparing to observational data -- the shattering hypothesis worked well at high redshift, but less well at low redshift.}. Later work on multi-phase gas in a turbulent medium has found a broad power-law spectrum of cloud sizes, with $dN/dM \propto M^{-2}$, or equal mass per logarithmic interval \citep{gronke2022,das23,singh2024}. At fixed density, this is equivalent to equal volume per logarithmic interval. Since $f_{\rm A} \propto f/r_c$, the smallest clouds will dominate the area covering factor, so a key parameter is the small-scale cutoff to this power-law spectrum. In \citet{gronke2022},  this small-scale cutoff is -- modulo uncertainty due to limited numerical resolution -- of order the survival radius (the minimal scale where cooling can overcome the mixing due to shear-induced Kelvin Helmholz instabilities), which is larger than the scale predicted by \citet{2018MNRAS.473.5407M}. There are many complications (e.g. non-thermal pressure support) that can affect these estimates, but it is fair to say that the lower limit on cool cloud sizes for $z \sim 0$ halos is very likely to be significantly larger than $\sim 1$~pc.

Increasing $r_c$ also alleviates the problem posed by \cite{jow2024}, who argue that refractive multiple imaging of FRBs by CGM clouds would suppress Galactic scintillation (and hence that detection of Galactic scintillation may disfavor some misty CGM models). Increasing $r_c$ from their nominal value of $r_c \sim 0.1$ pc to $r_c \sim 10$ pc decreases the number of refractive images by a factor of $10^{-10}$, to $N_{\rm image} \approx 1$ (see Equation 27 in \citealt{jow2024}). We thus find that VP19's CGM model can produce scattering levels comparable to our results if the model's cloud size is increased, and that under these same conditions refractive multiple imaging (and significant quenching of Galactic scintillation) is unlikely. {\cite{masribas2025} independently find that refractive multiple imaging is unlikely for typical cool CGM gas parameters. Ultimately, detailed assessment of the phase structure function behavior is likely needed to determine the best observational tests for distinguishing between these different scattering scenarios (refractive scattering by cool clouds vs. diffractive scattering by cool clouds and/or turbulence).}

The single cloud-scale model adopted by VP19 is qualitatively reminiscent of early monoscale descriptions of scintillation in both the interplanetary medium (IPM) and the ISM, in which density irregularities were modeled as a Gaussian power spectrum with a single characteristic length scale, as opposed to a turbulent power-law spectrum covering a wide range of scales \citep{lovelace70}. Under certain conditions (e.g., $\beta \geq 4$, or $\beta < 4$ and $l_{\rm i}\gtrsim r_{\rm diff}$), a power-law medium can produce a phase structure function with a similar dependence on spatial separation as a Gaussian density power spectrum (see \citealt{rickett77} for a review). Gaussian density spectra were largely ruled out based on observational evidence for density fluctuations spanning a much wider range of spatial scales than allowed in the monoscale model \citep{armstrong81b,armstrong95,2010ApJ...710..853C}. Distinguishing between monoscale and turbulent media in the CGM will similarly require assessment of observations spanning a wide range of spatial scales; measuring FRB scattering delays in a single radio frequency band will likely not suffice. However, the potential compatibility of non-thermal line broadening measurements with FRB scattering upper limits, as illustrated in this paper, already hints at fluctuations spanning a much wider range of scales than a monoscale model (such as VP19) would allow.

\section{Conclusions}\label{sec:conc}

{Quasar observations of CGM turbulence suggest that its density fluctuations are extremely weak: In the hot CGM, quasar-inferred Mach numbers and local electron densities yield $10^{-16} \lesssim C_{\rm n}^2 \lesssim 10^{-9}$ m$^{-20/3}$, whereas in the cool CGM, $10^{-8} \lesssim C_{\rm n}^2 \lesssim 10^{-4}$ m$^{-20/3}$. For typical path lengths $\sim 100$ kpc through a halo, the corresponding SM of the hot gas is $\sim 10^{-14} - 10^{-7}$ kpc m$^{-20/3}$. For the cool gas, a tiny volume filling fraction $f\sim10^{-3}-10^{-2}$ implies that the mean $\overline{C_{\rm n}^2}$ along the entire path length through a halo is $\sim 10^{-3} - 10^{-2}$ times smaller than the locally estimated values, and the resulting SM of the cool gas can be $1-10^3\times$ larger than that of the hot gas, depending on the volume-averaged density. The same result is obtained if one instead assumes that the locally estimated $C_{\rm n}^2$ of cool gas applies over a much smaller path length, such as that of an individual cool cloud. If the filling factor of cool gas is very small, then estimates of the mean $\overline{C_{\rm n}^2}$ along an FRB LOS, e.g. based on the inferred SM, are not necessarily capable of distinguishing scattering by cool vs. hot gas.} 

{The density fluctuation parameter $\tilde{F}$ will be substantially different in the cool and hot CGM due to the difference in their dissipation scales, filling factors, and clumpiness. If we assume that density fluctuations persist down to plasma microscales, then for the hot CGM we find a characteristic range $10^{-8} \lesssim \tilde{F} \lesssim 10^{-4}$ (pc$^2$ km)$^{-1/3}$, whereas for the cool CGM $10^{-4} \lesssim f\times\tilde{F} \lesssim 10^{-3}$ (pc$^2$ km)$^{-1/3}$. When the dissipation scale is significantly smaller than the diffractive scale, $l_{\rm i} \ll r_{\rm diff}$, the scattered pulse will show an extended tail with a mean delay that depends on the dissipation scale. Inferring $\tilde{F}$ from the mean scattering delay can thus serve as a sensitive diagnostic of the microphysics of the gas, if the scattering delay is actually measurable.}

{The range of $\tilde{F}$ found above implies that the expected level of diffractive scattering from the hot CGM is miniscule, but may be detectable from the cool CGM depending on the LOS geometry. Scattering delays from the Milky Way and host galaxy CGMs are predicted to be $\lesssim$~$\mu$s, regardless of gas phase. For a broad range of sightline geometries and likely DMs, we predict mean scattering delays from the hot gas $\langle \tau \rangle(1+z_\ell)^3 \ll 1\ \mu$s at 1 GHz, even after accounting for geometric amplification from intersecting a foreground CGM at cosmological distances. The cool CGM produces about $10^3\times$ larger scattering delays than the hot CGM, despite it being a smaller fraction of the total CGM electron density content. Even so, scattering from the cool CGM will likely only be detectable as pulse broadening at low observing frequencies, when the CGM contribution to DM and geometric leverage are very large, and/or if multiple halos are intersected and produce larger cumulative scattering (unless the ISMs of the Milky Way and host galaxy produce an inhibitive amount of scattering). It is also more likely that density fluctuations are damped in the hot gas at scales greater than the multipath scale, in which case the hot gas would not produce any multipath propagation. We thus conclude, similar to previous studies, that if FRB scattering occurs in the CGM, then cool gas will be the dominant contributor.}

{Our analysis indicates that even if CGM scattering is undetected, stringent upper limits on the mean scattering delay can be combined with quasar observations to yield lower limits on the dissipation scale. An initial comparison between quasar-inferred values of $\tilde{F}$ for FRB181112 and FRB191108 suggests that the scattering of these FRBs is plausibly dominated by their foreground CGMs, and that it is compatible with dissipation near plasma microscales. The dissipation scale can be more tightly contrained if a firm measurement of CGM scattering is made, but this scenario likely requires either negligible scattering from the ISMs of the FRB host galaxy and Milky Way, or highly robust determinations of their scattering contributions.}

{Importantly, we have shown that observed FRB scattering is not inconsistent with the misty CGM picture, as quasar observations of cool gas turbulence indicate small levels of scattering would occur even if that turbulence extends down to sub-au scales. We thus find that cool gas can be ubiquitous in the CGM without producing extreme amounts of FRB scattering incompatible with observations. Whether that cool gas is misty or concentrated in larger discrete clouds (more precisely, kinematically coherent structures) is difficult to assess. Scattering from a single discrete cloud corresponds to very large $G_{\rm scatt}$ because of the small path length through the cloud, which can yield significantly larger scattering than typically observed. One likely example of this is FRB20221219A, which intersects multiple foreground halos and has $\tau_{\rm d}\approx19$ ms at 1 GHz \citep{faber24}. Ultimately, targeted surveys of quasars and FRBs probing the same galaxy systems, with the requisite dedication of observing resources, offer the best prospects for constraining CGM structure across the full range of spatial scales necessary to test misty CGM models. The methodology pursued here is also applicable to FRB host galaxies, for which ISM turbulence may readily be studied in emission.}

\acknowledgements{The authors thank the anonymous referee, Cameron Hummels, Hsiao-Wen Chen, Gwen Rudie, Irina Zhuravleva, Jim Cordes, Drummond Fielding and James Beattie for insightful conversations about this work. PS thanks Pawan Kumar for his introduction to the basic physics of radio scintillation. Part of this work was performed at the Aspen Center for Physics, which is supported by National Science Foundation grant PHY-2210452. SKO and MC are supported by the Brinson Foundation through the Brinson Prize Fellowship Program. SKO is a member of the NANOGrav Physics Frontiers Center (NSF award PHY-2020265). SPO acknowledges NSF grant AST2407521 for support. Caltech and Carnegie Observatories are located on the traditional and unceded lands of the Tongva people.}

\appendix

\section{{Relating $C_{\rm n}^2$ and $\tilde{F}$}}
\label{app}

\begin{figure*}
    \centering
    \includegraphics[width=0.8\textwidth]{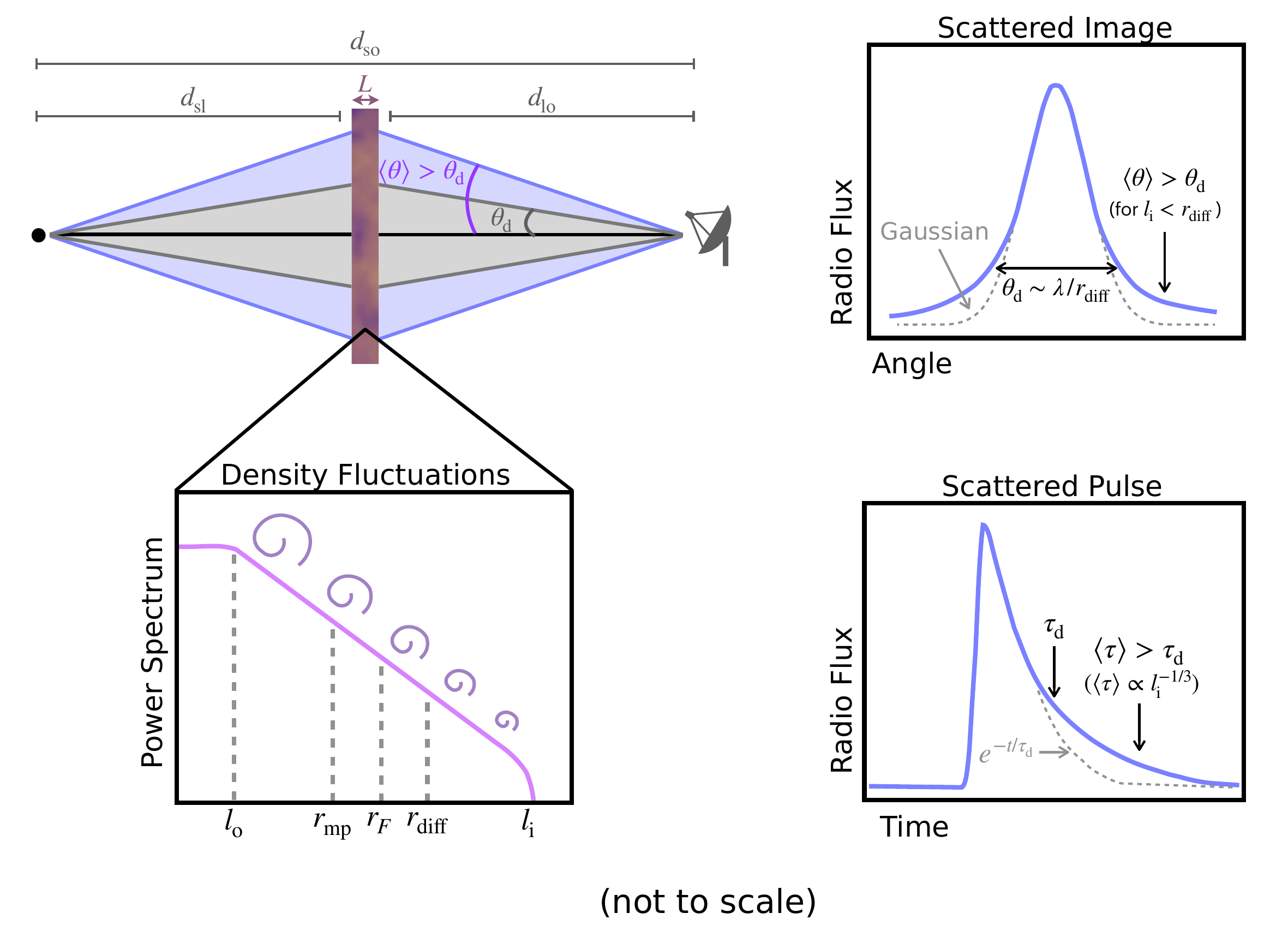}
    \caption{{Schematic illustration of the turbulent scattering screen (left) and scattered flux in the image and time domains (right), for the regime in which the inner scale is much smaller than the diffractive scale ($l_{\rm i} < r_{\rm diff}$). The diffractive scattering angle and delay are denoted by $\theta_{\rm d}, \tau_{\rm d}$, whereas the mean scattering angle and delay are denoted by $\langle \theta \rangle, \langle \tau \rangle$. The scattered image deviates from a Gaussian, and the scattered pulse has a tail that extends beyond the $1/e$ time (see Section~\ref{sec:obs-FRB} and Appendix~\ref{app}).}}
    \label{fig:scattering_diagram}
\end{figure*}

Here we lay out the formalism relating $C_{\rm n}^2$, the amplitude of the density fluctuation power spectrum, and $\tilde{F}$, the composite density fluctuation parameter in the ionized cloudlet model (Section~\ref{sec:obs-FRB}), expanding on the approach laid out in \cite{cordes2016}. We define the mean scattering time, {{i.e., the mean delay of the pulse broadening function,}} for a source at distance $d_{\rm so}$ 
\begin{equation}\label{eq:tau-eta}
   \langle \tau \rangle = (1/2c) \int_0^{d_{\rm so}} ds \eta(s) s(1-s/d_{\rm so})
\end{equation}
where $\eta(s)$ is the mean-square scattering angle per unit length due to turbulent plasma at a distance $s$:
\begin{equation}
    \eta(s) = \lambda^4 r_e^2 \int dq q^3 P_{\delta n_e}(q,s).
\end{equation}
Here {we have assumed an isotropic 3D density fluctuation spectrum} $P_{\delta n_e}$, and for a spectral index $\beta <4$ and an inner scale $l_{\rm i} = 2\pi/q_{\rm i}$ much smaller than the outer scale $l_{\rm o}$, the mean-square scattering angle is \citep{cordesrickett98}:
\begin{equation}\label{eq:mean-angle}
\begin{split}
    \eta(s) &\approx \frac{\lambda^4 r_e^2 \Gamma(3 - \beta/2)}{4-\beta} q_{\rm i}^{4-\beta} C_{\rm n}^2(s) \\
    &\approx \frac{1}{3}\lambda^4 r_e^2 \Gamma(7/6) q_{\rm i}^{1/3} C_{\rm n}^2(s)\ {\rm for} \ (\beta = 11/3).
\end{split}
\end{equation}
{{The mean-square scattering angle is distinct from the diffractive scattering angle ($\theta_{\rm d} \sim \lambda/r_{\rm diff}$). Unlike the diffractive scattering angle, the mean-square scattering angle will always depend on the inner scale and have the form given in Equation~\ref{eq:mean-angle} when $\beta < 4$. The corresponding mean delay (Equation~\ref{eq:tau-eta}) is not equal to the diffractive scattering delay (the delay corresponding to $\theta_{\rm d}$) when $l_{\rm i} \ll r_{\rm diff}$, as discussed in Section~\ref{sec:obs-FRB}.}} 

In the cloudlet model, individual cloudlets sustain internal turbulence with rms density fluctuations $\langle \delta n_{ec}^2 \rangle$ related to $C_{\rm n,c}^2$, where $C_{\rm n,c}^2$ refers to the turbulence amplitude within individual cloudlets:
\begin{equation}
\begin{split}
    C_{\rm n,c}^2 &= \frac{\beta-3}{2(2\pi)^{4-\beta}} \langle \delta n_{ec}^2 \rangle l_{\rm o}^{3-\beta} \\
    &= [3(2\pi)^{1/3}]^{-1} \langle \delta n_{ec}^2 \rangle l_{\rm o}^{-2/3}\ {\rm for} \ \beta=11/3.
\end{split}
\end{equation}
Letting $\epsilon^2 = \langle \delta n_{ec}^2 \rangle/n_{ec}^2$ and $C_{\rm SM} = (\beta-3)/2(2\pi)^{4-\beta}$, we thus have
\begin{equation}
    C_{\rm n,c}^2 = C_{\rm SM} \epsilon^2 n_{ec}^2 l_{\rm o}^{3-\beta},
\end{equation}
emphasizing that $n_{ec}$ is the local mean density of a cloudlet. Ultimately, we are interested in evaluating $\eta(s)$ for the volume-averaged $C_{\rm n}^2 = \overline{C_{\rm n,c}^2}$ along the LOS:
\begin{equation}
    \begin{split}
        \overline{C_{\rm n,c}^2} &= f C_{\rm SM} \langle \epsilon^2 n_{ec}^2\rangle l_{\rm o}^{3-\beta}\\
        &= f C_{\rm SM}  \langle \epsilon^2 n_{ec}^2\rangle l_{\rm o}^{3-\beta} \times \frac{\langle f n_{ec} \rangle^2}{\langle f n_{ec} \rangle^2} \\
        &= C_{\rm SM} \frac{ \zeta \epsilon^2}{fl_{\rm o}^{\beta-3}} \bar{n}_e^2,
    \end{split}
\end{equation}
where $\zeta = \langle n_{ec}^2 \rangle / \langle n_{ec} \rangle^2$ describes the fractional variations in density between cloudlets, $\bar{n}_e = f\langle n_{ec}\rangle$ is the volume-averaged electron density along the LOS, and we have assumed the outer scale and $\epsilon^2$ to be the same for all cloudlets.

Given the dependence of $\eta(s)$ on $q_{\rm i}$, we lump all quantities describing the density fluctuations together into the parameter $\tilde{F} = \zeta \epsilon^2/fl_{\rm o}^{\beta-3} l_{\rm i}^{4-\beta}$. We thus have 
\begin{equation}
\begin{split}
    \overline{C_{\rm n,c}^2} 
    &= C_{\rm SM} \tilde{F} \bar{n}_e^2 l_{\rm i}^{4-\beta} \\
    &= C_{\rm SM} \tilde{F} \bar{n}_e^2 l_{\rm i}^{1/3}\ {\rm for}\ \beta=11/3.
\end{split}
\end{equation}
The subsequent mean scattering time can be evaluated using Equation~\ref{eq:tau-eta}, as laid out in \cite{cordes2016,cordes2022}. For $\beta = 11/3$, this procedure ultimately yields the form given in Equation~\ref{eq:tau-Ftilde}, where we have used the screen dispersion measure due to a given phase ${\rm DM}_\ell = \bar{n}_e L$. The frequency dependence of $\eta(s)$, and hence the mean pulse broadening delay, is exactly $\nu^{-4}$, whereas the diffractive scattering delay has a frequency dependence that depends on $\beta$. Modifying $\beta$ would only change the constant prefactor in Equation~\ref{eq:tau-Ftilde} and the exponents of the inner and outer scales in $\tilde{F}$; all other quantities would remain the same. {{Ultimately, the mean scattering delay will always depend on $l_{\rm i}$ when $l_{\rm i} \ll r_{\rm diff}$, while $r_{\rm diff}$ behaves in the opposite manner -- $r_{\rm diff}$ only depends on $l_{\rm i}$ when $r_{\rm diff} \ll l_{\rm i}$}.}

\bibliography{master_bib}

\end{document}